\documentclass[11pt]{article}
\pdfoutput=1
\usepackage{amsmath}
\usepackage{amssymb}
\usepackage{graphicx,bbm,mathrsfs}
\usepackage{nicefrac}
\usepackage{slashed}
\usepackage{bbm}
\usepackage{geometry}
\usepackage{empheq}
\usepackage{stackrel}
\usepackage{ulem}
\usepackage{jheppub}
\usepackage{setspace}
\usepackage{relsize}
\usepackage{empheq}
\usepackage{pifont}
\usepackage{mathtools}
\usepackage{enumitem}
\usepackage{dcolumn}   
\usepackage{bm}        
\usepackage{graphicx,mathrsfs}
\usepackage{nicefrac}
\usepackage{multirow}
\usepackage{color}
\usepackage{mathtools}
\usepackage{makecell}
\usepackage{environ} 
\usepackage{lipsum} 
\usepackage{wasysym}
\usepackage{hhline,colortbl}  
\usepackage{tikz}
\usepackage{graphicx}
\usepackage{tensor}
\usepackage{dsfont}
\usepackage{simpler-wick}
\usepackage{accents}

\usepackage{booktabs}
\usepackage{mdframed}
\usepackage{xcolor}
\definecolor{EqFrame}{RGB}{235,245 ,250 }

 \NewEnviron{Smaller11}{
           \scalebox{1.1}{$\BODY$} 
 } 
 \NewEnviron{Smaller08}{
           \scalebox{0.8}{$\BODY$} 
 }

\newcommand{\be}{\begin{equation}}
\newcommand{\ee}{\end{equation}}
\newcommand{\bea}{\begin{eqnarray}}
\newcommand{\eea}{\end{eqnarray}}

\renewcommand\labelenumi{(\roman{enumi})}
\renewcommand\theenumi\labelenumi

\usepackage{titlesec}

\titleformat*{\section}{\Large\bfseries}
\titleformat*{\subsection}{\large\bfseries}
\titleformat*{\subsubsection}{\large\bfseries}
\titleformat*{\paragraph}{\large\bfseries}
\titleformat*{\subparagraph}{\large\bfseries}

\makeatletter
\newcommand*{\prodsym}{%
  \DOTSB
  \mathop{
    \mathchoice
      {\rlap{\kern.3em\rotatebox[origin=c]{-90}{}}{\prod}}
      {\vcenter{\rlap{\kern.2em\rotatebox[origin=c]{-90}{}}}{\prod}}
      {\sum}{\sum}
  }\slimits@
}
\makeatother

\makeatletter
\DeclareFontFamily{OMX}{MnSymbolE}{}
\DeclareSymbolFont{MnLargeSymbols}{OMX}{MnSymbolE}{m}{n}
\SetSymbolFont{MnLargeSymbols}{bold}{OMX}{MnSymbolE}{b}{n}
\DeclareFontShape{OMX}{MnSymbolE}{m}{n}{
    <-6>  MnSymbolE5
   <6-7>  MnSymbolE6
   <7-8>  MnSymbolE7
   <8-9>  MnSymbolE8
   <9-10> MnSymbolE9
  <10-12> MnSymbolE10
  <12->   MnSymbolE12
}{}
\DeclareFontShape{OMX}{MnSymbolE}{b}{n}{
    <-6>  MnSymbolE-Bold5
   <6-7>  MnSymbolE-Bold6
   <7-8>  MnSymbolE-Bold7
   <8-9>  MnSymbolE-Bold8
   <9-10> MnSymbolE-Bold9
  <10-12> MnSymbolE-Bold10
  <12->   MnSymbolE-Bold12
}{}

\let\llangle\@undefined
\let\rrangle\@undefined
\DeclareMathDelimiter{\llangle}{\mathopen}%
                     {MnLargeSymbols}{'164}{MnLargeSymbols}{'164}
\DeclareMathDelimiter{\rrangle}{\mathclose}%
                     {MnLargeSymbols}{'171}{MnLargeSymbols}{'171}
\makeatother

\begin{document}

\vspace*{4mm}

\thispagestyle{empty}

\begin{center}

\begin{center}
\huge
{\bf Upper Bounds on the Radion Mass \\ in a Warped Extra Dimension}
\end{center}

\vspace{2cm}

\renewcommand{\thefootnote}{\fnsymbol{footnote}}

{\large\bf
Leandro~Da Rold$^{\, a}$\,\footnote{daroldl@ib.edu.ar}\,, 
Eugenio~Meg\'{\i}as$^{\, b}$\,\footnote{emegias@ugr.es}\,,
Mariano~Quir\'os$^{\, c}$\,\footnote{quiros@ifae.es}\,,
Lindber~Salas$^{\, a}$\,\footnote{lindber.salas@ib.edu.ar}
}

\vspace{0.5cm}
\noindent

${}^a\!$ 
\textit{Centro At\'omico Bariloche, Instituto Balseiro and CONICET,} \\
\indent \; \textit{Av. Bustillo 9500, 8400, S.C. de Bariloche, Argentina}

${}^b\!$ 
\textit{Departamento de F\'{\i}sica At\'omica, Molecular y Nuclear and} \\
\indent \; \textit{Instituto Carlos I de F\'{\i}sica Te\'orica y Computacional,} \\
\indent \; \textit{Universidad de Granada, Avenida de Fuente Nueva s/n, 18071 Granada, Spain}

${}^c\!$  
\textit{Institut de F\'{\i}sica d'Altes Energies (IFAE) and} \\
\indent \; \textit{The Barcelona Institute of  Science and Technology (BIST),} \\
\indent \; \textit{Campus UAB, 08193 Bellaterra, Barcelona, Spain}

\end{center}


\vspace*{8mm}
 
\begin{center}
  {  \bf  Abstract }
  
\end{center}
\begin{minipage}{15cm}
\setstretch{0.95}

We study absolute upper bounds on the mass ratio $R_{11}=m_{r_1}/m_{h_1}$ between the radion ($r_1$) and graviton ($h_1$) lightest modes in five dimensional warped models. We consider warped models with  
an admissible singular behavior in the infrared: i.e.~models with power-like ($\sim \bar\phi^{2n}$) and with exponential ($\sim e^{2\nu\bar\phi}$) bulk potentials, where $\bar\phi$ is the stabilizing bulk field. 
We find a correlation between $R_{11}$ and the singular behavior of the Ricci scalar, which allows to establish an absolute upper bound on the former. 
For the power-like models, the maximum value of $R_{11}$ increases with $n$. In the limit $n\to\infty$, the power-like model reproduces the exponential model and reaches the same upper bound as the latter. For the exponential model, we derive analytically the bound $R_{11}<1$, with the largest value approached for the case $\nu\to 1^+$ (where $\nu=1$ corresponds to the linear dilaton model) and consistent with unitarity sum rules. For exponential models we also obtain the absolute upper bound on the second radion mode  ratio $R_{21}=m_{r_2}/m_{h_1}\lesssim 1.374$. This value is reached for $\nu=2$ when the IR brane approaches the singularity. For this configuration, we verify that the scalar contribution satisfies the unitarity sum rule when the complete radion Kaluza-Klein tower is included.

\vspace{0.5cm}
\end{minipage}

\setcounter{tocdepth}{2}
\setcounter{footnote}{0}

\tableofcontents  

\vspace{1cm}
\hrule
\vspace{1cm}

\newpage

\section{Introduction}
\label{sec:introduction}
The existence of curled up extra dimensions has been proposed as an extension of our four-dimensional universe in order to solve some of the well known problems of the Standard Model of electroweak and strong interactions. Such extra dimensions appear e.g.~as a mean to unify gravity with the other known interactions in the ten-dimensional supersymmetric strings, with Planck size extra dimensions (for a recent review see e.g.~Ref.~\cite{Marchesano:2024gul}), or to solve the hierarchy problem, as the solution proposed by Arkani-Hamed, Dimopoulos and Dvali~\cite{Arkani-Hamed:1998jmv,Antoniadis:1998ig}, which explains why gravity is so much weaker than the other fundamental forces by introducing  a number $n\geq 2$ of (relatively) large flat extra spatial dimensions, such that the higher dimensional Planck scale is at the TeV and the hierarchy is explained geometrically by the volume of the extra dimensions. A particularly appealing solution to the hierarchy problem was proposed by Randall and Sundrum (RS)~\cite{Randall:1999ee,Randall:1999vf} who introduced a warped extra dimension, with two four dimensional branes, at the Planck [ultraviolet (UV) brane] and TeV [infrared (IR) brane] scales respectively, such that the hierarchy between them is explained by the warp factor of the five-dimensional (5D) geometry. A common feature to all these theories is the presence of Kaluza-Klein (KK) gravitons, with mass $m_{h_n}$, which appear as excitations of the higher dimensional metric tensor when projected onto the four space-time dimensions.

The presence of scalar fields fixing the geometry of the internal manifold $\mathcal M$, called moduli, is a general feature of extra-dimensional theories. Moduli are typically massless until they are provided by some dynamics lifting the corresponding flat directions, giving them a mass and fixing the shape and size of the internal manifold. When moduli are massless, the internal manifold can deform without any energetic costing. On the other hand, the more massive moduli are, the more costly are internal manifold deformations. In other words, rigidity $\mathcal R$ of the internal manifold against deformations is tied to the mass of moduli fields, so a simple definition of rigidity can be given as $\mathcal R(\mathcal M)=m_{\ell}^2/m_{h_1}^2$ where $m_{\ell}$ is the lightest physical geometric modulus mass. Given thus an extra dimensional theory with an internal manifold $\mathcal M$, knowledge of its rigidity amounts to knowledge of the moduli mass spectrum, and how rigid a manifold can be amounts to provide an upper bound on $m_\ell$.

There has recently been some activity in investigating how rigid an internal manifold can be, using the four dimensional effective field theory of moduli and the first graviton KK mode $h_1^{\mu\nu}$, obtained after integrating out all other KK modes, by imposing perturbative unitarity of the $h_1 h_1\to h_1 h_1$ scattering amplitude~\cite{Mirbabayi:2026saz,Bertucci:2026osd}~\footnote{The unitarity of 
the RS theory in the presence of the GW stabilization mechanism has also been studied in Refs.~\cite{Chivukula:2021xod,Chivukula:2022tla,Chivukula:2023sua,Chivukula:2023qrt}.}. The common result in Refs.~\cite{Mirbabayi:2026saz,Bertucci:2026osd} is that the maximal rigidity is provided by the upper bound on $m_\ell$ given by
\be
\frac{m_\ell^2}{m_{h_1}^2}\leq \frac{4}{3}  \,,
\label{eq:bound-unitarity}
\ee
where strict equality is discarded as it would imply the vanishing of all cubic couplings involving one modulus and two KK gravitons $h_1^{\mu\nu}h_1^{\alpha\beta}$.

In this paper we will attack this problem for the case of a general warped five dimensional theory, using five dimensional techniques to compute the spectra. As the theory is five dimensional the internal manifold is a circle compactified on the $S^1/\mathbb Z_2$ orbifold, and as such the theory contains a single modulus, the radion. We will assume the radion is stabilized by the Goldberger-Wise (GW) mechanism~\cite{Goldberger:1999uk} which provides a mass for the radion $m_r$ and fixes the brane distance. The GW mechanism uses a bulk scalar field $\bar\phi$ with (stiff) brane potentials fixing its VEV at each brane. In the absence of the GW mechanism the radion is massless, and when there is a weak back-reaction the radion is light, such that $m_r\ll m_{h_1}$, so by going to strong back-reaction regimes we expect to get heavy radions and thus to reach the rigidity limits of the internal manifold. In particular we will explore consistent theories with a naked singularity in the metric~\cite{Gubser:2000nd,Cabrer:2009we}, i.e.~theories exhibiting a Ricci scalar with a singularity for $\bar\phi\to\infty$ at $y=y_s$ ($y$ being the proper coordinate), where the back-reaction on the metric is the strongest one~\footnote{Notice that a naked singularity is acceptable only if it can be interpreted as the zero-temperature limit of a regular black-hole solution (more precisely, if the singularity can be cloaked by a regular horizon at finite temperature).}.
Then although, admittedly, there is no direct relation between rigidity and singular behavior of the Ricci scalar, we will find the heaviest radion (maximal rigidity) for the theory where the Ricci scalar has the strongest singularity. Interestingly enough, and anticipating the final result, the latest theory turns out to have the linear dilaton background~\cite{Antoniadis:2001sw,Antoniadis:2011qw,Megias:2021mgj,Fichet:2023xbu,Megias:2026wwp}, an  effective theory with a stringy dual, the Little String Theory~\cite{Aharony:1999ks,Aharony:1999ti}.

As we are looking after theories with a strong IR deformation we will use the superpotential method $W(\bar\phi)$~\cite{DeWolfe:1999cp} to carefully take into account the back-reaction on the metric, and seek in the catalog of theories exhibited in Ref.~\cite{Cabrer:2009we}, classified according to their behavior at $\bar\phi\to\infty$ with an increasing strength of the singularity. 

The first class of theories are those with polynomial superpotential, i.e.~$W(\bar\phi)\sim \bar\phi^n$.  For $n\leq 2$ there is no singularity, unless of course when $y_s\to \infty$. For $n>2$ the theories have a singularity at a finite value of $y_s$, and $z_s\to\infty$, where $z$ is the conformal coordinate. We have proven that the critical behavior of the Ricci scalar increases with $n$ as $R\sim (\log z)^n$ and the rigidity $\mathcal R(n)$ also increases with $n$. In the limit $n\to\infty$ the rigidity reaches the result from exponential superpotentials.

The second class of theories then consists of exponential superpotentials as $W(\bar\phi)\sim e^{\nu\bar\phi}$, with $\nu$ a positive real number. This class of models reproduces AdS for $\nu=0$ and the linear dilaton background for $\nu= 1$. In the absence of IR brane, this class of models then exhibits a metric singularity at $z=z_s$ and the Ricci scalar shows a critical behavior as $R\sim z^\gamma$ with a critical exponent $\gamma={2\nu^2/|\nu^2-1|}$ which has a singularity at $\nu\to 1$, while the rigidity $\mathcal R(\nu)$ also has a maximum, not unexpectedly, at $\nu\to 1^+$. In fact for the linear dilaton case $\nu=1$ the Ricci scalar behaves exponentially $R\sim e^{2 z/y_s}$ which grows faster with $z$ than any power.  For $\nu<2$ the naked singularities are acceptable, however they are not for $\nu\geq 2$. So the cases where $W(\bar\phi)>e^{2\bar\phi}$ will not be considered any longer as they would correspond to naked singularities which are not shielded by horizons.

The maximum value of the rigidity for the considered warped model is then the corresponding value for $\nu\to 1^+$, which has been found to be
\be
\frac{m_{r_1}^2}{m_{h_1}^2} < 1 \,,
\label{eq:R11}
\ee
where $r_1$ is the lightest radion mode,
which is consistent with the general bound (\ref{eq:bound-unitarity}) obtained from unitarity in the effective theory in Ref.~\cite{Mirbabayi:2026saz}, assuming discrete spectra. Notice that our result includes the case of spectra with a gapped continuum, as the case of the $\nu=1$ model where the IR brane can be replaced by the singularity, or more precisely by the location of the horizon shielding the singularity. 

Models with light radions have been previously explored in the literature with power-like potentials, leading to radion masses $m_{r_1}\ll m_{h_1}$~\cite{Bellazzini:2013fga,Megias:2014iwa,Pomarol:2019aae,Elander:2020fmv,Pomarol:2023xcc,Elander:2025fpk,Piai:2026rst}. In particular, a light radion can persist even in regimes of strong backreaction~\cite{Pomarol:2019aae,Pomarol:2023xcc}. This shows that strong backreaction alone does not necessarily lead to a heavy radion. In the models with a singularity considered here, instead, the presence of the singularity allows the lightest radion to become as heavy as the first graviton.

Moreover, for models with exponential superpotential, we have also found the upper bound on the second radion mode
\be
\frac{m^2_{r_2}}{m^2_{h_1}}\leq\left(\frac{x_2}{x_1}\right)^2=\left(\frac{5.266}{3.832} \right)^2\simeq 1.89  \,,
\label{eq:R21}
\ee
where $m_{r_2}$ is the mass of the next to lightest radion KK mode $r_2$, and $x_{1,2}$ are the first solutions to $J_1(x_1)=0$ and $x_2 J_2(x_2)=(2/3)J_1(x_2)$, respectively. Actually, the upper bound in Eq.~(\ref{eq:R21}), achieved for the case $\nu=2$ and $z_1=z_s$,
exceeds the unitarity bound of Eq.~(\ref{eq:bound-unitarity}) although there is no contradiction as the theory is mainly unitarized by the lightest modulus, as we will see. 

We have then argued that the degree of rigidity of the theory follows the strength of the divergent behavior of its Ricci scalar in conformal coordinates. This seems to be a general pattern although we do not have a formal proof of it. We expect to go back further on this point in a future publication.

The contents of this paper are as follows. In Sec.~\ref{sec:model} we will introduce our formalism for arbitrary 5D models with general metric $A$, general bulk potentials, and general scalar potentials localized in the branes. We also introduce the superpotential formalism to deal with the backreaction of the scalar field on the gravitational metric. In Sec.~\ref{sec:explicit_realizations} we apply the general formalism to two particularly interesting classes of models exhibiting good naked singularities: models with power-law and models with exponential potentials. For the previous models, which  exhaust those with an asymptotic behavior exhibiting good naked singularities, we have observed a correlation between the strength of the singularity and the degree of rigidity of the 5D manifold. In Sec.~\ref{sec:ratios} we have worked out numerically the ratios $m_{r_i}/m_{h_1}$, for $i=1,2$, of the first two radion mode masses with respect to the first graviton mode mass, in both power-like and exponential models. We have found that the bound in Eq.~(\ref{eq:R11}), the upper bound on the corresponding ratio, corresponds to the case of the exponential model with $\nu\to1^+$, where the model with $\nu=1$ is the linear dilaton model, in the case where the IR brane overlaps the singularity. Moreover, in the exponential model we have also found that the inequality (\ref{eq:R21}) is saturated in the case of the exponential model for the limiting case $\nu=2$ when the IR brane overlaps the singularity. In Sec.~\ref{sec:maxR11} we provide a formal analytical derivation of the bound (\ref{eq:R11}) on $m_{r_1}/m_{h_1}$ in the exponential model. In Sec.~\ref{sec:unitarity} we show that the upper bounds we have found for $m_{r_1}/m_{h_1}$ and $m_{r_2}/m_{h_1}$ do not violate the unitarity constraints for the scattering amplitude $\mathcal M(h_1 h_1 \to h_1 h_1)$. Moreover for the exponential model where the IR brane overlaps the singularity we have also found that the infinite tower giving rise to the mass ratios $m_{r_\ell}/m_{h_1}$ does not either violate the unitarity constraints. More general results on unitarity are postponed to future research. Our results and outlook can be found in Sec.~\ref{sec:conclusions}. Finally, more technical details about different stages of the calculation can be found in Apps.~\ref{Ap:KK} and~\ref{Ap:unitarity}. 
%


\section{The five dimensional model}
\label{sec:model}

We follow the notation of Refs.~\cite{Fichet:2023xbu,Barbosa:2024pyn}. The scalar-gravity system is given by the action
\begin{equation}
  S = \int d^5x \sqrt{-g} \left( \frac{M_5^3}{2} R - \frac{1}{2} (\partial_M \phi)^2 - V(\phi)\right) - \sum_{\alpha} \int d^4x \sqrt{-g} \left( V_\alpha(\phi) - M_5^3 K_\alpha \right) \,,
\end{equation}
where $R$ is the 5D Ricci scalar, $\phi$ the scalar field, $M_5$ the fundamental 5D Planck scale $(\kappa^2 = 1 / M_{5}^3)$, and the indices $\alpha = 0\,(1)$ stand for the UV (IR) brane. The brane $\alpha$ supports a brane localized potential $V_\alpha(\phi)$. $K_\alpha$ is the extrinsic curvature of brane $\alpha$ that appears in the Gibbons-Hawking-York (GHY) boundary term. The $V_\alpha(\phi)$ potential stabilizes $\phi$ to the vacuum expectation value (VEV) $\langle \phi\rangle_\alpha \equiv v_\alpha$, which determines completely the background.

\subsection{The background}

Let us consider a metric in proper coordinates of the form
\begin{equation}
ds^2 = e^{-2A(y)} \eta_{\mu\nu} dx^\mu dx^\nu + dy^2 \,,
\end{equation}
where the 4D Minkowski metric is $\eta_{\mu\nu} = \textrm{diag}(-1,1,1,1)$. The on-shell action can be written in the form
\begin{equation}
S_{\textrm{on-shell}} = - \int d^4x  \, V_{\textrm{eff}}(y_0,y_1) \,, 
\end{equation}
where
\begin{equation}
  V_{\textrm{eff}}(y_0,y_1) = e^{-4A(y_1)} U_{1}(\phi_1) - e^{-4A(y_0)} U_{0}(\phi_0)
\end{equation}
is the radion effective potential, and 
\begin{equation}
U_\alpha(\phi) \equiv V_\alpha(\phi) - (-1)^\alpha W(\phi)
\end{equation}
are localized effective potentials, while $W$ is the superpotential, which is solution of the~EoM (\ref{eq:VW}) written below. For convenience, we will define the dimensionless scalar field $\bar\phi$ as
\begin{equation}
\bar\phi \equiv \frac{\phi}{\sqrt{3} M_5^{3/2}} \,,
\end{equation}
as well as the rescaled potentials and superpotential
\begin{equation}
\bar V \equiv \frac{V}{3 M_5^3} \,, \qquad  \bar V_\alpha \equiv \frac{V_\alpha}{3 M_5^3} \,,\qquad \bar U_\alpha \equiv \frac{U_\alpha}{ 3 M_5^3} \,,  \qquad  \bar W \equiv \frac{W}{3 M_5^3} \,.
\end{equation}
These quantities have mass dimensions $[\bar V] = M^2$ and $[\bar V_\alpha] = [\bar U_\alpha] =  [\bar W] = M$.  

Using this notation, the background equations of motion (EoM) are given by
\begin{eqnarray}
  A^{\prime\prime}(y) &=& \bar\phi^{\prime \, 2}(y) + \sum_\alpha \bar V_\alpha(\bar\phi) \delta(y - y_\alpha) \,, \\
  A^{\prime \, 2}(y) &=& - \frac{1}{2} \bar V(\bar \phi) + \frac{1}{4} \bar\phi^{\prime\, 2}(y) \,, \\
  \bar\phi^{\prime\prime}(y) - 4 A^\prime(y) \bar\phi^\prime(y) &=& \bar V^\prime(\bar \phi) + \sum_\alpha \bar V_\alpha^\prime(\bar \phi) \delta(y - y_\alpha) \,.
\end{eqnarray} 
In terms of the superpotential, the EoM's write 
\begin{equation}
A^\prime(y) = \frac{1}{2} \bar W(\bar \phi) \,, \qquad \bar \phi^\prime(y) = \frac{1}{2} \frac{\partial \bar W( \bar \phi)}{\partial \bar\phi} \,,  \label{eq:EoM_background}
\end{equation}
and
\begin{equation}
  \bar V(\bar\phi) = \frac{1}{8}  \left( \frac{\partial \bar W(\bar\phi)}{\partial \bar\phi} \right)^2 -  \frac{1}{2} \bar W(\bar \phi)^2   \,. \label{eq:VW}
\end{equation}
Moreover, the 5D Ricci scalar can be written as
\begin{equation}
R = 8 A^{\prime\prime}(y) - 20 A^\prime(y)^2 \,. \label{eq:R_5D}
\end{equation}

\subsection{The radion/scalar system}
\label{sec:radion}

The radion follows from a scalar perturbation of the metric of the form
\begin{eqnarray}
  ds^2 &=& e^{-2A(y) - 2 F(x,y)} \eta_{\mu\nu} dx^\mu dx^\nu + [1 + G(x,y)]^2 dy^2 \,, \\
  \bar\phi(x,y) &=& \bar\phi(y) + \bar\varphi(x,y) \,,
\end{eqnarray}
where the three scalars $F$, $G$ and $\bar\varphi$ are not independent functions but satisfy the relations
\begin{align}
  F(x,y) &= F(y)  \cdot \mathcal R(x) \,, \qquad \bar\varphi(x,y) = \bar\varphi(y) \cdot \mathcal R(x) \,,  \label{eq:Fxy}\\
 \bar\phi^\prime(y) \bar\varphi(y)  &= F^\prime(y) - 2 A^\prime(y) F(y) \,, \qquad G(x,y) = 2 F(x,y) \,, \label{eq:varphiy}
\end{align}
where we indicate the mode expansion as: $A(y) \cdot B(x) \equiv \sum_n A^{(n)}(y) B^{(n)}(x)$. This means that we are assuming a factorization $A^{(n)}(y) \, B^{(n)}(x)$ for each mode $n$, so that the relations of Eqs.~(\ref{eq:Fxy}) and (\ref{eq:varphiy}) are valid both for any mode $n$, and for the summation of all the modes. As we can see from Eq.~(\ref{eq:Fxy}) the field $\mathcal R(x)$ is the excitation of a linear combination of the radion $F(x,y)$ and the scalar $\bar\phi(x,y)$ perturbations with different weights which, from Eq.~(\ref{eq:varphiy}), depend on the profile $F(y)$. So all along this paper we will indistinctly name $\mathcal R(x)$ as radion or scalar. 

Using the background EoM one can recast the bulk EoM and boundary condition for the excitation $F$, with mass $m_r$, as~\cite{Megias:2015ory}
\begin{eqnarray}
\label{eq-KK-radion-1}
  0 &=& \partial_y\left(  \frac{e^{2A}}{A^{\prime\prime}}  \partial_y\left[ e^{-2A} F \right] \right) + \left( m_r^2 \frac{e^{2A}}{A^{\prime\prime}}  - 2 \right)F  \,, \\
  0 &=& \left[ (-1)^\alpha m_r^2 F + \bar U_\alpha^{\prime\prime} \partial_y\left( e^{-2A} F \right) \right]_{y_\alpha}  \,.
\end{eqnarray}
As one of the main objectives of this work is to study configurations leading to a heavy radion, in the following we will work in the stiff wall limit in both branes, i.e. $\bar U^{\prime\prime}_\alpha \to \infty \; (\alpha = 0,1)$, by which the boundary conditions reduce to
\be
\left.\partial_y\left(e^{-2A} F\right)\right|_{y_\alpha}=0\,.
\ee
Based on an analytical analysis from mass formulae and on numerical computations, it was argued in e.g.~Ref.~\cite{Megias:2015ory} that this limit maximizes the radion mass.

\subsection{The graviton}
\label{sec:graviton}

The graviton follows from the fluctuation of the metric of the form
\begin{equation}
ds^2 = e^{-2A(y)} \left( \eta_{\mu\nu}  + \frac{2}{M_5^{3/2}}  h_{\mu\nu}(x,y) \right) dx^\mu dx^\nu + dy^2 \,.
\end{equation}
By using the ansatz
\begin{equation}
h_{\mu\nu}(x,y) = h_{\mu\nu}(x)  h(y) \,,
\end{equation}
the EoM for the profile $h(y)$ corresponding to an excitation of mass $m_h$ is given by
\begin{equation}
 0 = m_h^2 h + e^{2A}\partial_y\left( e^{-4A} h^\prime \right)  \,,  \label{eq:EoM_fraviton}
\end{equation}
with boundary conditions
\begin{equation}
h^\prime(y) \big|_{y_0, y_1}  = 0  \,.
\end{equation}

\subsection{Schr\"odinger form of the KK equations}
\label{sec:schrod}
By using conformal coordinates $z$, given by $dz=e^{A}dy$, the KK equations of the radion and graviton, (\ref{eq-KK-radion-1}) and (\ref{eq:EoM_fraviton}), can be written in Schr\"odinger form:
   \begin{eqnarray}
-\tilde F^{\prime\prime}(z) + V_F(z)\tilde F(z)
&=& m_r^2 \tilde F(z) \,,
\qquad
F(z)=e^{3A(z)/2}\bar\phi^\prime(z)\tilde F(z)  \,,
\label{eq:Sch_F}
\\
-\tilde h^{\prime\prime}(z) + V_h(z)\tilde h(z)
&=& m_h^2 \tilde h(z) \,,
\qquad \; h(z)=e^{3A(z)/2}\tilde h(z) \,,
\label{eq:Sch_h}
\end{eqnarray}
  where the effective Schr\"odinger potentials are
  \begin{eqnarray}
    V_F(z) &=&  \frac{9}{4} A^\prime(z)^ 2 + \frac{5}{2} A^{\prime\prime}(z) - \frac{\bar\phi^{\prime\prime}(z)}{\bar\phi^\prime(z)} A^\prime(z) + 2 \left( \frac{\bar\phi^{\prime\prime}(z)}{\bar\phi^\prime(z)} \right)^2 - \frac{\bar\phi^{\prime\prime\prime}(z)}{\bar\phi^\prime(z)}  \,, \label{eq:VF} \\
    V_h(z) &=& \frac{9}{4} A^\prime(z)^2 - \frac{3}{2} A^{\prime\prime}(z) \,, \label{eq:Vh}
  \end{eqnarray}
  while the corresponding boundary conditions at the branes for these fields are
  \begin{eqnarray}
   \partial_z\left( e^{-2A(z)} F(z) \right) \Big|_{z_0, z_1} = 0 \,, \qquad \partial_z  h(z) \big|_{z_0, z_1}  = 0 \,.   \label{eq:BC_radion_graviton}
  \end{eqnarray}

\section{Explicit realizations of the model}
\label{sec:explicit_realizations}

We consider two classes of warped models, generated by power-like and exponential superpotentials. We first present the corresponding backgrounds and Schr\"odinger potentials, and then discuss the limit in which both models converge, before comparing the corresponding mass ratios.

\subsection{The power-like potential}
\label{subsec:power}

Let us consider a superpotential of the form
\begin{equation}
\bar W(\bar\phi) = 2k \bar\phi^n \,, \qquad \forall n \,,
\end{equation}
with the corresponding bulk potential given by
\begin{equation}
\bar V(\bar\phi) = 2 k^2 \bar\phi^{2(n-1)} \times \left[ \left( \frac{n}{2} \right)^2 - \bar\phi^2 \right]\,.
\end{equation}
The solution of the EoM's of the background~(\ref{eq:EoM_background}) is given by
\begin{eqnarray}
  A(y) &=& \frac{n-2}{2} k y_s \bar\phi_0^n \times \left[ \left(  1 - \frac{y}{y_s} \right)^{\frac{2}{2-n}} - 1 \right] \,, \label{eq:Ay_n} \\
  \bar\phi(y) &=& \bar \phi_0  \times \left(  1 - \frac{y}{y_s} \right)^{\frac{1}{2-n}} \,, \label{eq:phiy_n}
\end{eqnarray}
where the value of the scalar field at $y=0$ is given in terms of the singularity location $y_s$ by 
\begin{equation}
\bar\phi_0 = \left( n(n-2) k y_s \right)^{\frac{1}{2-n}}  \,, \label{eq:phi0}
\end{equation}
and we will find it convenient to introduce a real parameter $\nu$ by
\be
\bar\phi_0\equiv \frac{n}{\nu}  \,,
\ee
as we will make connection with the class of models with exponential superpotentials studied in Sec.~\ref{subsec:exponential}. From this equation, one finds that the position of the singularity in proper coordinates is at
\begin{equation}
y_s = \frac{\bar\phi_0^{2-n}}{n(n-2) k} \,. \label{eq:relation-ys}
\end{equation}

We will focus on values $n > 2$. By using Eqs.~(\ref{eq:phiy_n}) and~(\ref{eq:phi0}), we can also write the solution for the warp factor as
\begin{equation}
A(y) = \frac{1}{2 n} \bar\phi_0^2 \times \left[ \left(  1 - \frac{y}{y_s} \right)^{\frac{2}{2-n}} - 1 \right]  = \frac{1}{2n} \left( \bar\phi^2 - \bar\phi_0^2 \right) \,.  \label{eq:Ayphi}
\end{equation}
From this expression, it is clear that the masses of the radion and graviton cannot depend directly on $k$, but on the parameters $\{ \bar\phi_0, n, y_1, y_s \}$. For fixed values of $\bar\phi_0$ and $n$, the masses scale as $m \propto k$, or equivalently $m \propto 1/y_s$.

The relation between proper and conformal coordinates is
\begin{equation}
z(y) = -\frac{(n-2)}{2} y_s \, e^{-\alpha} \times  \left\{  \left(1 - \frac{y}{y_s} \right)  \times E_{n/2}\left[ - \alpha \left( 1 - \frac{y}{y_s} \right)^{\frac{2}{2-n}} \right] - E_{n/2}\left[ - \alpha \right]  \right\} \,, \label{eq:zy_n}
\end{equation}
where
\begin{equation}
\alpha \equiv \frac{1}{2n} \bar\phi_0^2 =\frac{n}{2\nu^2} \,,
\end{equation}
and $E_n(z) = \int_1^\infty dt \frac{e^{-zt}}{t^n}$ is the exponential integral function. In this expression we have chosen $z_0 \equiv z(y=0) = 0$, a convention we will adopt throughout the remainder of this manuscript. The approximate formula near the singularity $(y \simeq y_s)$ is
%
%
%
\begin{equation}
z(y) \simeq  \frac{y_s(n-2)\nu^2}{n} \left(1 - \frac{y}{y_s} \right)^{\frac{n}{n-2}} \exp\left[ \frac{n}{2\nu^2}  \, \left(1 - \frac{y}{y_s} \right)^{-\frac{2}{n-2}} -1\right]  \,. \label{eq:zy_powerlike}
\end{equation}
The solution to Eq.~(\ref{eq:zy_powerlike}) for $n$ finite is given by
\be
1-\frac{y}{y_s}=\left[-\nu^2 \mathcal W_{-1}\left( -\frac{e^{-1/\nu^2}}{n^2 (kz)^{2/n}}\right) 
  \right]^{-(n-2)/2}  \,, 
\ee
and the argument of the Lambert function goes to zero as $z\to\infty$, so that $1-y/y_s\to 0$ for $n>2$. More in detail
\be
1-\frac{y}{y_s}\simeq \left(\frac{n}{2\nu^2 \log(kz)} \right)^{(n-2)/2}\,,
\ee
and it is clear that $\lim_{y \to y_s} z(y) = +\infty$ for $n > 2$. 

On the other hand, the Ricci scalar is
  %
  %
   \begin{equation}
  R = -4k^2 \bar\phi_0^{2n} \times \left[ 5  \left( 1 - \frac{y}{y_s} \right)^{-\frac{2n}{n - 2}} - 2\nu^2 \left( 1 - \frac{y}{y_s} \right)^{- \frac{2(n-1)}{n - 2}} \right]  \,, \label{eq:R_powerlike}
  \end{equation}
and in the limit $z\to\infty$, and $n$ finite, one gets
  \be
R\simeq -4k^2\bar\phi_0^{2n}\left(\frac{2\nu^2 \log(kz)}{n} \right)^n\left[5-\frac{n}{\log(kz)} \right]\,.
  \ee
  
For finite $n>2$, the large-$z$ behavior of the Ricci scalar is therefore controlled by $[\log(kz)]^n$. Away from this asymptotic region, the exact relation between proper and conformal coordinates is given by Eq.~(\ref{eq:zy_n}), which cannot in general be inverted in elementary form at finite $n$. It is therefore more convenient to express the Schr\"odinger potentials directly in terms of $\bar\phi$. Using Eq.~(\ref{eq:Ayphi}), the potentials defined in Eqs.~(\ref{eq:VF}) and~(\ref{eq:Vh}) can be written as
    \begin{eqnarray}
V_h(\bar\phi) &=& \frac{3}{4} k^2  e^{ - \frac{\bar\phi^2 - \bar\phi_0^2 }{ n } }  \bar\phi^{2(n-1)}  \left( 5 \bar\phi^2 - 2 n^2 \right) \,, \\
V_F(\bar\phi) &=&  \frac{1}{4} k^2   e^{ - \frac{\bar\phi^2 - \bar\phi_0^2 }{ n } } \bar\phi^{2(n-2)}  \left[ 3 \bar\phi^4 + 2 n (4 + 3 n) \bar\phi^2 + 4 n^2 (n-1) \right] \,.
    \end{eqnarray}

\subsection{The exponential potential}
\label{subsec:exponential}
Let us now consider the exponential model, for which the background can be written explicitly both in proper and conformal coordinates. The superpotential and potential are given by
%
\begin{equation}
\bar W(\bar\phi) = 2 k \, e^{\nu \bar\phi}  \,, \qquad \bar V(\bar\phi) = -\frac{k^2}{2} (4 - \nu^2) \, e^{2 \nu \bar\phi} \,.
\end{equation}
The solution of the EoM of the background in proper coordinates is given by
\begin{equation}
A(y) = - \frac{1}{\nu^2} \log\left( 1 - \frac{y}{y_s} \right) \,,
\qquad
\bar\phi(y) = -\frac{1}{\nu} \log\left[\nu^2 k (y_s - y)\right] \,,
\label{eq:exp_background_y}
\end{equation}
where the singularity location $y_s$ is, in principle, an arbitrary parameter.

For $\nu\neq 1$, in the conformal coordinate introduced in Sec.~\ref{sec:schrod}, the solution is
\begin{eqnarray}
  A(z) &=& \frac{1}{1-\nu^2} \log\left[ 1 - \bar\rho (z - z_{0}) \right]  \,, \label{eq:Aphi_L1} \\
  \bar\phi(z) &=&   \frac{\nu}{1 - \nu^2} \log\left[ 1 - \bar\rho ( z - z_{0} ) \right]  - \frac{1}{\nu} \log\left( \nu^2 \frac{k}{\rho} \right) \,,  \label{eq:Aphi_L2}
\end{eqnarray}
where we have conveniently defined
\be
\bar\rho \equiv \frac{(\nu^2 - 1)}{\nu^2} \rho \,,
\ee
and, from a comparison of the value of the scalar field $\bar\phi$ at the UV brane in proper and conformal coordinates, one finds
\begin{equation}
\rho = \frac{1}{y_s} \,.
\end{equation}

The relation between proper and conformal coordinates is
\begin{equation}
\rho (z - z_0 )  = \frac{\nu^2}{(1-\nu^2)} \left[  \left( 1 - \frac{y}{y_s} \right)^{\frac{\nu^2 - 1}{\nu^2}} - 1 \right] \,.
\end{equation}
In these expressions, $z_{0}$ is the position of the UV brane which can be chosen arbitrarily. In the following we will assume $z_0 = 0$, without loss of generality.

The Ricci scalar can be written as
\begin{equation}
R = 4 (2\nu^2 - 5)\rho^2
\left(1-\bar\rho z\right)^{\frac{2\nu^2}{1-\nu^2}} \, .
\label{eq:R_exp}
\end{equation}
The singularity is at $z \to z_s$ with
\begin{equation}
  z_s = \left\{ \begin{array}{cc}
    \infty & \quad (\nu < 1) \\
     1/\bar\rho  & \quad (\nu > 1)
     \end{array} \right. \,.  \label{eq:zs}
\end{equation}
Then, we can define the theory in two different intervals according with the classification of Ref.~\cite{Fichet:2023dju}: the interval $\mathcal M_\nu^-$ (where the massless graviton exists on top of the tower of KK gravitons) and $\mathcal M_\nu^+ $ (where only KK modes of the graviton propagate)
\begin{equation}
    \mathcal M^-_\nu = \left\{ \begin{array}{cc}
      \left[0, +\infty\right) & \quad (\nu < 1) \\
      \left[0, 1/\bar\rho\right) & \quad (\nu > 1)
    \end{array} \right. \,, \qquad
          \mathcal M^+_\nu = \left\{ \begin{array}{cc}
    (  1/\tilde\rho , 0] & \quad (\nu < 1) \\
      \left(-\infty, 0\right] & \quad (\nu > 1)
    \end{array} \right. \,.
\end{equation}
In this paper we will only be concerned by the physical interval $\mathcal M_\nu^-$, which always contains the singularity.

The graviton and radion potentials entering in the Schr\"odinger like KK Eqs.~(\ref{eq:Sch_F}) and~(\ref{eq:Sch_h}) are:
\begin{eqnarray}
V_h(z) &=& \frac{3}{4} (5 - 2 \nu^2) \rho^2
\frac{1}{(1 - \bar\rho z)^2} \,,
\label{eq:Vh_exp}
\\
V_F(z) &=& \frac{3}{4} (1 + 2 \nu^2) \rho^2
\frac{1}{(1 - \bar\rho z)^2} \,.
\label{eq:VF_exp}
\end{eqnarray}
The location of the singularity in conformal coordinates, cf.~Eq.~(\ref{eq:zs}), leads to a restriction in the domain of the parameter space in which the exponential model with two branes is well defined, leading to
\begin{equation}
   \nu < \nu_{\textrm{max}} = \left( 1 - \frac{1}{\rho z_1 } \right)^{-1/2} \,.
\end{equation}

The special case $\nu=1$ corresponds to the linear dilaton model. The well-defined $\nu\to1^\pm$ limit of Eqs.~(\ref{eq:Aphi_L1})-(\ref{eq:Aphi_L2}) is
\begin{equation}
  A(z) \to \rho \, z  \,, \qquad  \bar\phi(z) \to \rho \, z - \log\left( \frac{k}{\rho} \right) \,,  \quad  (\nu \to 1^\pm)   \,.\label{eq:linear_dilaton_background}
\end{equation}
In this case, the continuous limit $\nu\to1^\pm$ of Eq.~(\ref{eq:R_exp}) gives
\begin{equation}
R=-12\rho^2 e^{2\rho z} \, .
\label{eq:R_linear_dilaton}
\end{equation}
The singularity is located at $z_s=\infty$, and the physical interval is $\mathcal M^-_{\nu=1}=[0,+\infty)$. The Schr\"odinger potentials reduce to

\begin{equation}
V_h(z)=V_F(z)=\left(\frac{3\rho}{2}\right)^2. \,
\label{eq:V_linear_dilaton}
\end{equation}

Let us point out that, following Gubser's criterion~\cite{Gubser:2000nd}, the singularity of the exponential model is a good singularity for $\nu \in [0,2)$, whereas it is a bad singularity for $\nu \in [2,\infty)$, see e.g. Refs.~\cite{Cabrer:2009we,Fichet:2023dju,Barbosa:2024pyn}. This justifies restricting our analysis in the following to the parameter range $\nu \in [0,2)$ that yields a good singularity. Although the existence of a bad singularity is, in principle, irrelevant in the two-brane version of the model for which $\mathcal M^-_\nu = [0,z_1]$, it is convenient to restrict ourselves to the good singularity case since we are mostly concerned with the limit $z_1 \to z_s$.

\subsection{The $n\to\infty$ limit}
\label{subsec:n_infty}
We now return to the power-like model and take $n\to\infty$ with $\bar\phi_0=n/\nu$ and fixed $\nu$. At fixed $y/y_s$, Eqs.~(\ref{eq:Ay_n}) and~(\ref{eq:phiy_n}) give
\begin{equation}
A(y)\longrightarrow -\frac{1}{\nu^2}\log\left(1-\frac{y}{y_s}\right)\,,
\qquad
\bar\phi(y) \longrightarrow \bar\phi_0 - \frac{1}{\nu}\log\left(1-\frac{y}{y_s}\right)\,.
\label{eq:ninf_background_y}
\end{equation}
Since $\bar\phi_0=\bar\phi(0)$, these are precisely the background profiles of the exponential model in Eq.~(\ref{eq:exp_background_y}). Thus the power-like background tends to the exponential background as $n\to\infty$. In the conformal coordinate introduced in Sec.~\ref{sec:schrod}, the background in the limit $n \to \infty$ is correspondingly described by Eqs.~(\ref{eq:Aphi_L1}) and~(\ref{eq:Aphi_L2}).

Since the radion and graviton Schr\"odinger equations are fixed by the background, the KK spectrum of the power-like model is expected to tend to that of the exponential model as $n\to\infty$. The numerical results below confirm this behavior for the mass ratios. In particular, $m_{r_1}/m_{h_1}$ tends to the same upper bound, whose derivation for the exponential model is given in Sec.~\ref{sec:maxR11}.

\section{Results for the radion to graviton modes mass ratios}
\label{sec:ratios}
Motivated by the discussion of internal-manifold rigidity in Sec.~\ref{sec:introduction}, we define the ratio of scalar to graviton mode masses as:
\begin{equation}
R_{jk}\equiv\frac{m_{r_j}}{m_{h_k}} \ .
\end{equation}

The KK masses can be computed by making use of the equations of Secs.~\ref{sec:radion}-\ref{sec:schrod}, with the corresponding potentials to each model. For the power-like model, it is not possible in general to find an analytical solution of the KK equations; for that reason we rely on numerical methods for the calculation of $R_{jk}$ on that model. For the exponential model one can use Eqs.~(\ref{eq:graviton_spectral_sec3}) and (\ref{eq:scalar_spectral_sec3}), expressed in terms of Bessel functions, with the masses obtained numerically from them.

\subsection{The power-like model}
\begin{figure}[htb]
\begin{center}
\includegraphics[width=0.6\textwidth]{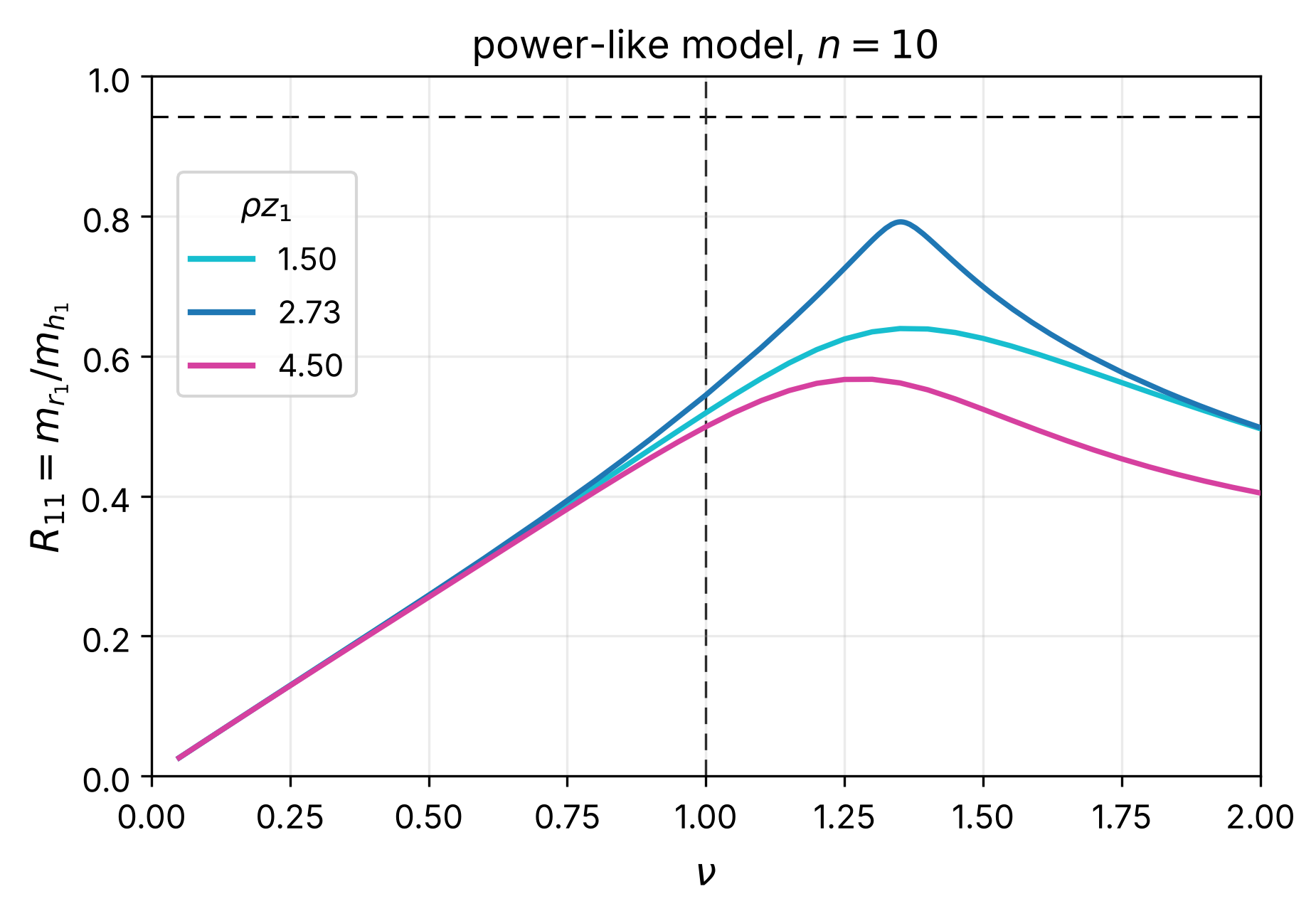}
\end{center}
\caption{\it $R_{11}=m_{r_1}/m_{h_1}$ as function of $\nu$ in the power-like model with $n=10$, for three values of $\rho z_1$.
The intermediate value, $\rho z_1=2.73$, is the one that contains the absolute maximum of $R_{11}$; the other two ($\rho z_1=1.50$ and $4.50$)
give lower maxima. The vertical dashed line marks $\nu=1$ and the horizontal one $2\sqrt{2}/3\simeq0.943$.}
\label{R11_n10_three}
\end{figure}

We start with the results for $R_{j1}$ in the power-like model. For each $n$, $R_{11}=m_{r_1}/m_{h_1}$ is a function of the two parameters $\nu$ and $\rho z_1$, and it has a maximum at a particular pair of values, which we locate numerically. Figure~\ref{R11_n10_three}
illustrates this for $n=10$. The value of $\rho z_1$ at the maximum is first obtained from the
numerical search; $R_{11}$ is then computed as a function of $\nu$ at that fixed value of $\rho z_1$.
For comparison, Fig.~\ref{R11_n10_three} also shows the same function for a smaller and a larger value of $\rho z_1$. The curve through the maximum ($\rho z_1\simeq2.73$, with $R_{11}^{\max}\simeq0.79$)
lies above the other two, whose maxima are lower. Both the height and the position of the
maximum depend on $\rho z_1$, while for small $\nu$ the three curves coincide and only
separate for larger $\nu$.

\begin{figure}[htb]
\begin{center}
\includegraphics[width=0.99\textwidth]{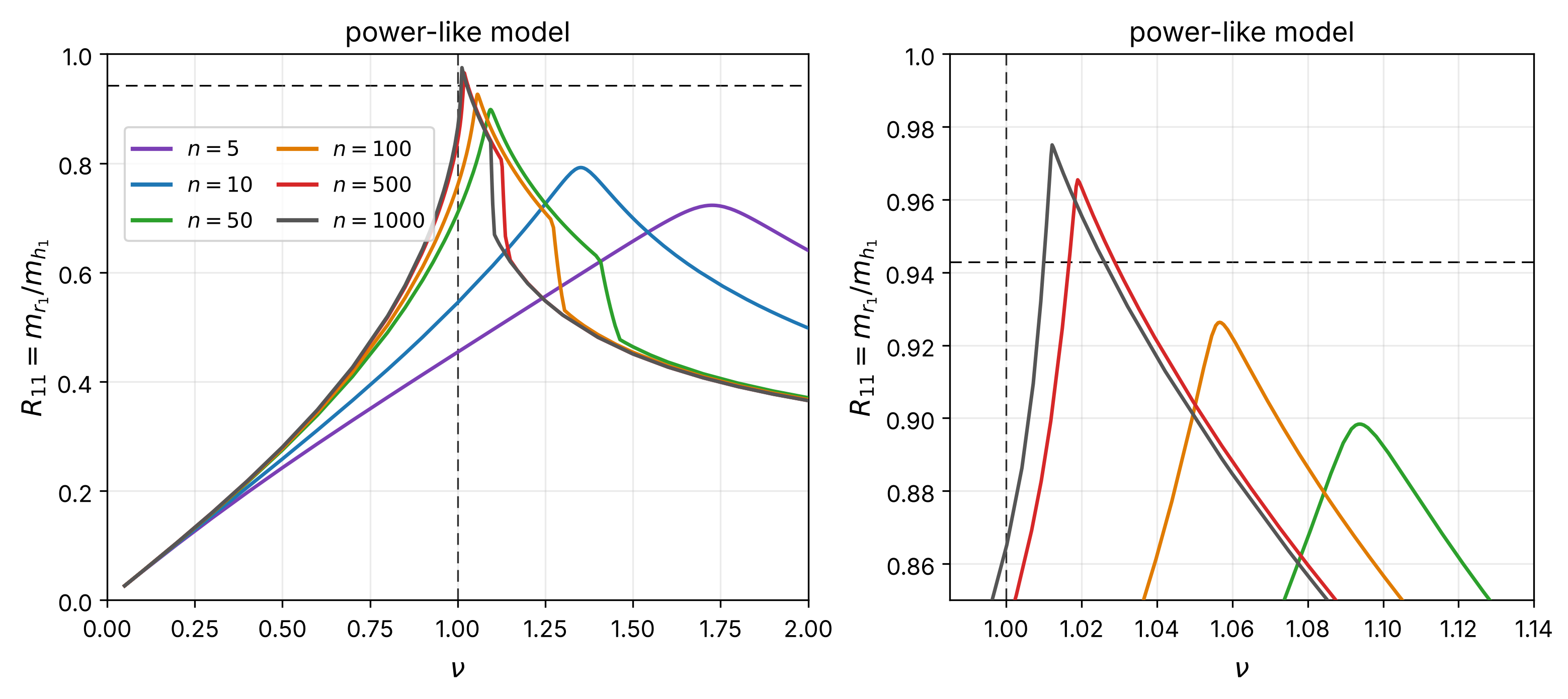}
\end{center}
\caption{\it$R_{11}=m_{r_1}/m_{h_1}$ as function of $\nu$ for different values of $n$,
in the power-like model. The values $n=5,\,10,\,50,\,100,\,500,\,1000$ correspond to $\rho z_1=1.95,\,2.73,\,4.87,\,6.09,\,10.09,\,12.59$, respectively. Left panel: full range of $\nu$. Right
panel: zoom around the maxima. The vertical dashed line marks $\nu=1$ and the horizontal one
$R_{11}=2\sqrt{2}/3\simeq0.943$.}
\label{R11_vs_nu_modeln}
\end{figure}

Figure~\ref{R11_vs_nu_modeln} shows $R_{11}$ as a function of $\nu$ for $n=5,\,10,\,50,\,100,\,500$ and $1000$, choosing the value of $\rho z_1$ that contains the maximum $R_{11}$ for that $n$, namely $\rho z_1=1.95$, $2.73$, $4.87$, $6.09$, $10.09$ and $12.59$, respectively. 
The sharp drops close to the right end of the curves for $n\geq50$ are generated by the crossing with heavier modes.

$R_{11}$ vanishes as $\nu\to0$ and, at numerical level, we find that the curves for different $n$
nearly coincide for $\nu\lesssim0.2$, and that they grow linearly.
As $\nu$ increases, $R_{11}$ grows monotonically up to a maximum whose height and position
depend on $n$: the maximum increases from $0.723$ at $\nu=1.726$ for $n=5$ to $0.975$ at
$\nu=1.012$ for $n=1000$. As $n\to\infty$, in agreement with Sec.~\ref{subsec:n_infty}, the power-like model tends to the exponential model
(see Fig.~\ref{R11_vs_nu_modelnu}), and $R_{11}$ tends to the same values: $R_{11}^{\max}\to1$ and
its position approaches $\nu=1$ (see Sec.~\ref{sec:maxR11}), while for $\nu<1$ the maximum approaches $2\sqrt{2}/3$ (see App.~\ref{app:mass_ratio_limits}).

\begin{figure}[htb]
\begin{center}
\includegraphics[width=0.99\textwidth]{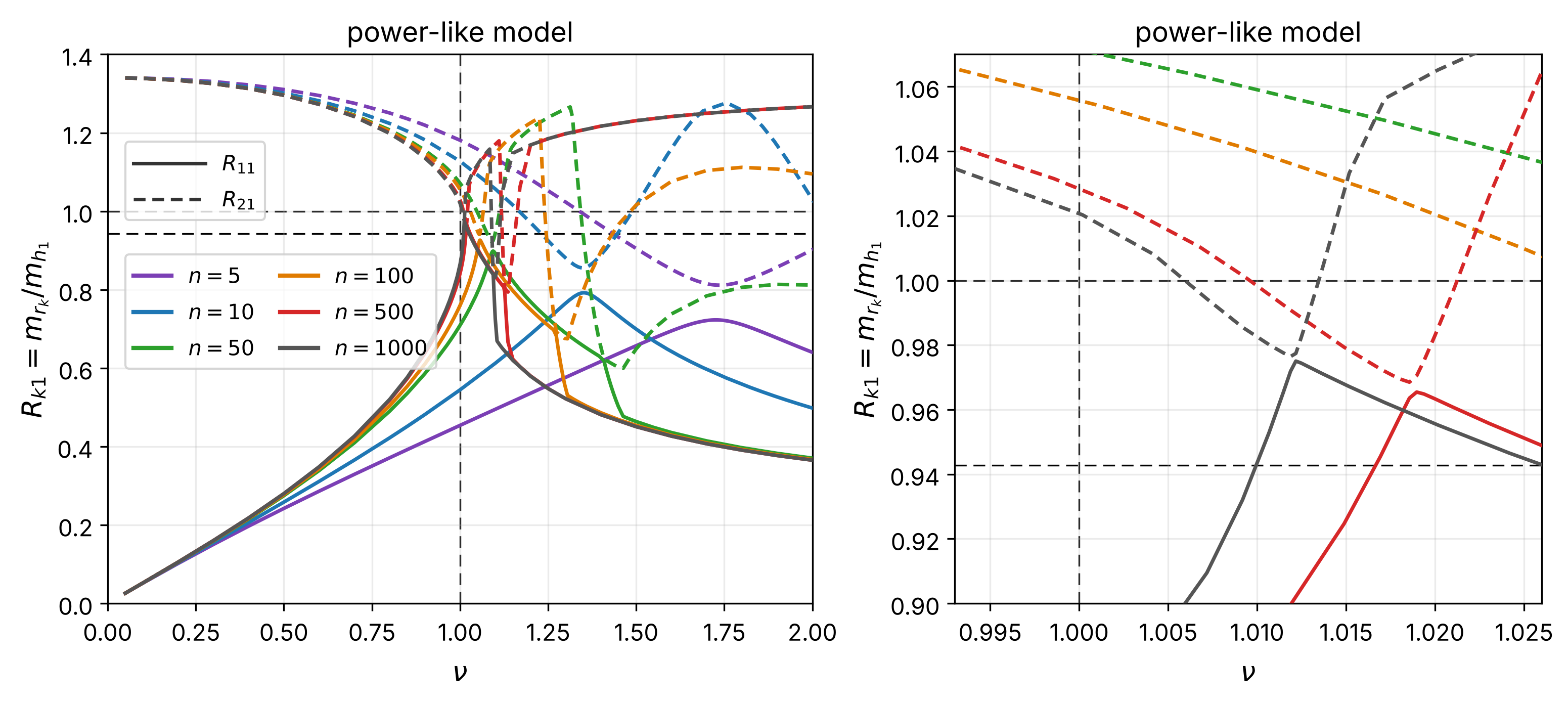}
\end{center}
\caption{\it $R_{11}=m_{r_1}/m_{h_1}$ (solid) and $R_{21}=m_{r_2}/m_{h_1}$ (dashed) as functions
of $\nu$ for different values of $n$, in the power-like model. The values $n=5,\,10,\,50,\,100,\,500,\,1000$ correspond to $\rho z_1=1.95,\,2.73,\,4.87,\,6.09,\,10.09,\,12.59$, respectively. Left panel: full range of $\nu$. Right panel: zoom around the maxima of $R_{11}$. The vertical dashed line marks $\nu=1$ and the
horizontal one $2\sqrt{2}/3\simeq0.943$.}
\label{R11_R21_vs_nu_modeln}
\end{figure}

Figure~\ref{R11_R21_vs_nu_modeln} shows $R_{21}=m_{r_2}/m_{h_1}$, the ratio for the second
scalar mode, together with $R_{11}$ in
the power-like model, for the same values of $n$ and $\rho z_1$ as in Fig.~\ref{R11_vs_nu_modeln}, i.e.,
$n=5,\,10,\,50,\,100,\,500$ and $1000$ and the corresponding $\rho z_1$. $R_{21}\to1.34$ for $\nu\to0$
for all $n$, whereas in contrast $R_{11}\to0$, and $R_{21}$ decreases as $\nu$ increases.
As can be seen in the right panel $R_{21}>1$ at $\nu=1$, approaching 1 as $n$ grows.

$R_{21}$ has a minimum at nearly the same value of $\nu$ as the maximum of $R_{11}$ and the difference
$R_{21}-R_{11}$ decreases with $n$. 
For $n=1000$ the two lightest scalar modes are almost degenerate there, since $m_{r_2}$ and $m_{r_1}$ differ by only $0.2\%$.
For larger $\nu$, $R_{21}$ increases up to a local maximum, which is followed by a decrease
of $R_{21}$ that becomes more abrupt as $n$ grows. For $n\geq50$ it is an abrupt drop, close
to the sharp drop of $R_{11}$ near the right end of the curves, while for $n=5$ and $10$ the
decrease is gradual and extends beyond the range of the figure.

Notice that for each $n$, $\rho z_1$ was chosen to maximize $R_{11}$. This choice does not guarantee that $R_{21}$ is also maximized; in fact we have checked numerically that, generically, $R_{21}$ is maximized at different values of the parameters.

\subsection{The exponential model}
\begin{figure}[htb]
\begin{center}
\includegraphics[width=0.99\textwidth]{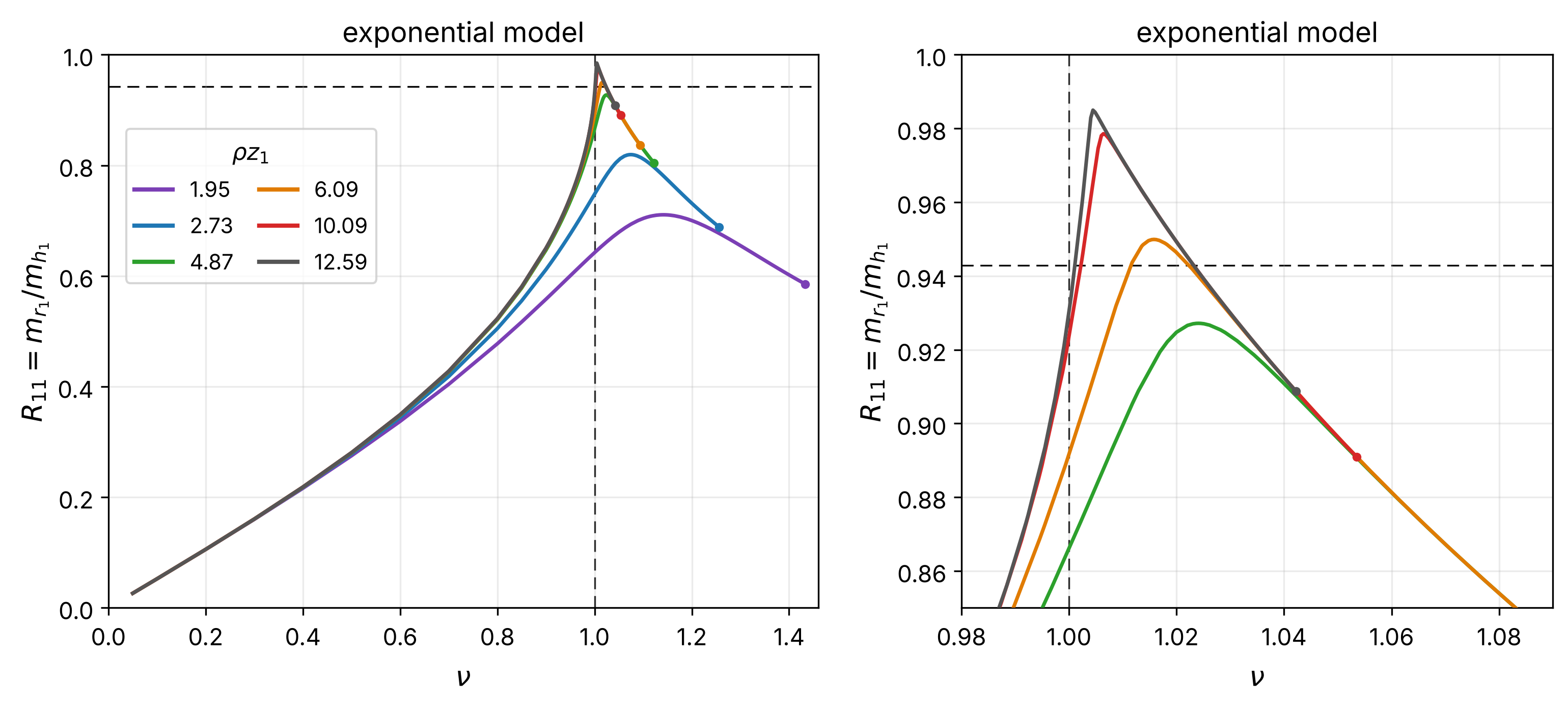} 
\end{center}
\caption{\it $R_{11}=m_{r_1}/m_{h_1}$ as function of $\nu$ for different values of $\rho z_1$, in the exponential model. Left panel: full range of $\nu$. Right panel: zoom around the maxima. Each curve ends (dot) at $\nu_{\max}$. The vertical dashed line marks $\nu=1$ and the horizontal one $R_{11}=2\sqrt{2}/3\simeq0.943$}
\label{R11_vs_nu_modelnu}
\end{figure}

Let us now consider the exponential model. Figure~\ref{R11_vs_nu_modelnu} shows $R_{11}$ as a function of $\nu$. For illustrative purposes, we take  the same fixed values of $\rho z_1$ as in the power-like model, namely $\rho z_1=1.95$, $2.73$, $4.87$, $6.09$, $10.09$ and $12.59$. Each curve is defined for $\nu<\nu_{\max}$ and ends (dots in the plot) at that
value, where the IR brane approaches the singularity. At the endpoint $R_{11}$ takes a
finite value: $0.585$, $0.688$, $0.805$, $0.837$, $0.891$ and $0.909$ for increasing
$\rho z_1$.
$R_{11}$ vanishes as $\nu\to0$, and at numerical level we find that the curves for different $\rho z_1$ nearly coincide for $\nu\lesssim0.4$, and that they grow linearly.
As $\nu$ increases, $R_{11}$ grows monotonically up to a maximum whose height and position depend on $\rho z_1$: the maximum increases from $0.711$ at $\nu=1.142$ for $\rho z_1=1.95$ to $0.985$ at $\nu=1.0045$ for
$\rho z_1=12.59$. Thus, as $\rho z_1$ grows, $R_{11}^{\max}\to1$ and its position approaches
$\nu=1$. As can be seen in the right panel, for $\nu<1$ the maximum approaches $2\sqrt{2}/3$ (horizontal dashed line). In Sec.~\ref{sec:maxR11} we will show analytically how $R_{11}$ reaches its maximum value.

\begin{figure}[htb]
\begin{center}
\includegraphics[width=0.99\textwidth]{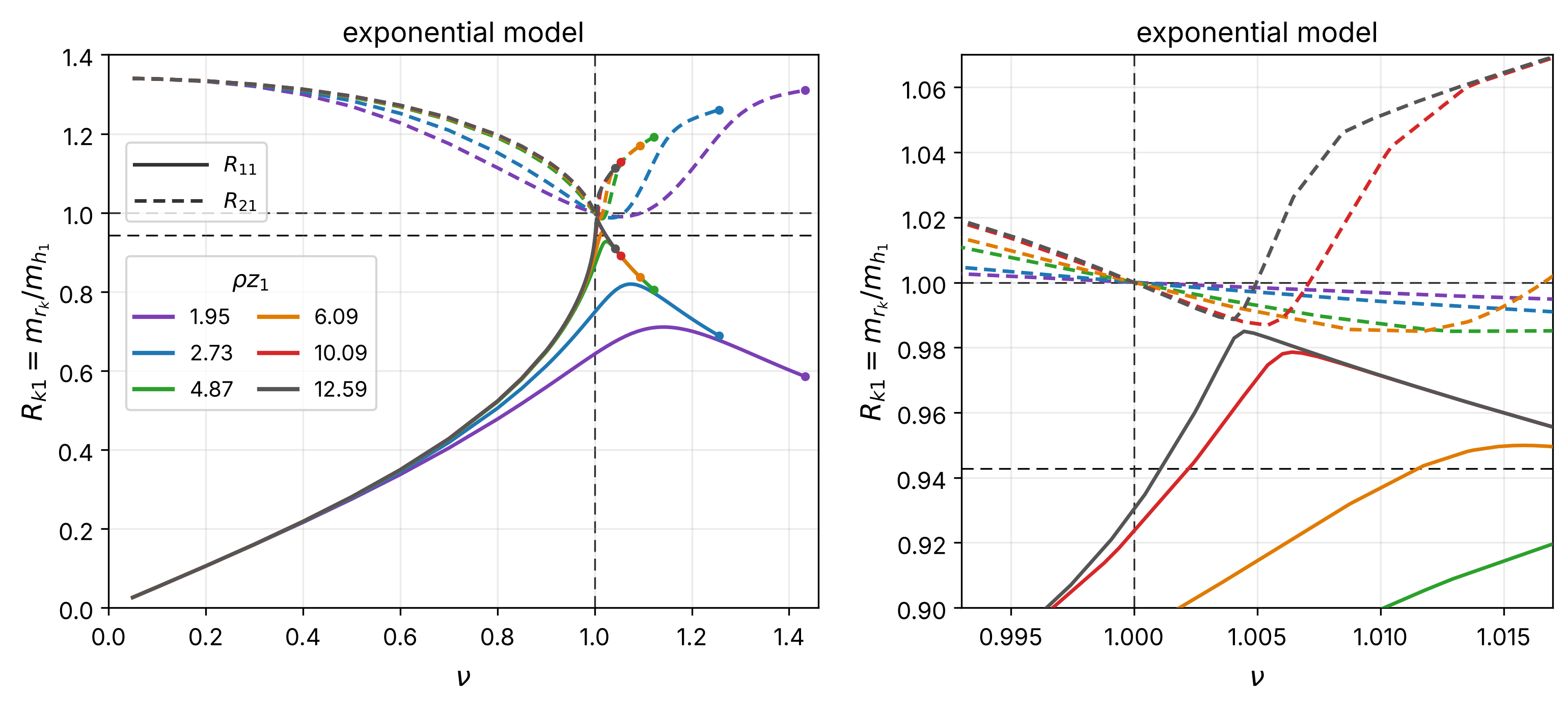}
\end{center}
\caption{\it $R_{11}=m_{r_1}/m_{h_1}$ (solid) and $R_{21}=m_{r_2}/m_{h_1}$ (dashed) as functions
of $\nu$ for different values of $\rho z_1$, in the exponential model. Left panel: full range
of $\nu$. Right panel: zoom around the maxima of $R_{11}$. Each curve ends (dot) at
$\nu_{\max}$. The vertical dashed line marks $\nu=1$ and the horizontal one
$2\sqrt{2}/3\simeq0.943$.}
\label{R11_R21_vs_nu_modelnu}
\end{figure}

Figure~\ref{R11_R21_vs_nu_modelnu} shows $R_{21}=m_{r_2}/m_{h_1}$ together with $R_{11}$, for the same values of $\rho z_1$. In contrast to
$R_{11}$, which vanishes as $\nu\to0$, $R_{21}\to1.34$ as in the power-like model, for
all values of $\rho z_1$. As $\nu$ increases, $R_{21}$ decreases from this value down to a
minimum, $R_{21}\simeq0.985$--$0.990$, slightly below 1, and, as can be seen in the right
panel, all the $R_{21}$ curves cross $R_{21}=1$ at $\nu=1$. The position of the minimum
depends on $\rho z_1$ and, like that of the maximum of $R_{11}$, approaches $\nu=1$ as
$\rho z_1$ grows. After the minimum $R_{21}$ increases, steeply for the largest values of
$\rho z_1$ and more gradually for the smallest ones.

Comparing both ratios, $R_{21}>R_{11}$ for all $\nu$, and the two curves come closest
around the maximum of $R_{11}$. The difference $R_{21}-R_{11}$ at that maximum decreases
as $\rho z_1$ grows.
For $\rho z_1=12.59$ the two lightest scalar modes are almost degenerate there:
$R_{11}=0.985$ and $R_{21}=0.992$, so $m_{r_2}$ and $m_{r_1}$ differ by only $0.7\%$.

\begin{figure}[htb]
\begin{center}
\includegraphics[width=0.49\textwidth]{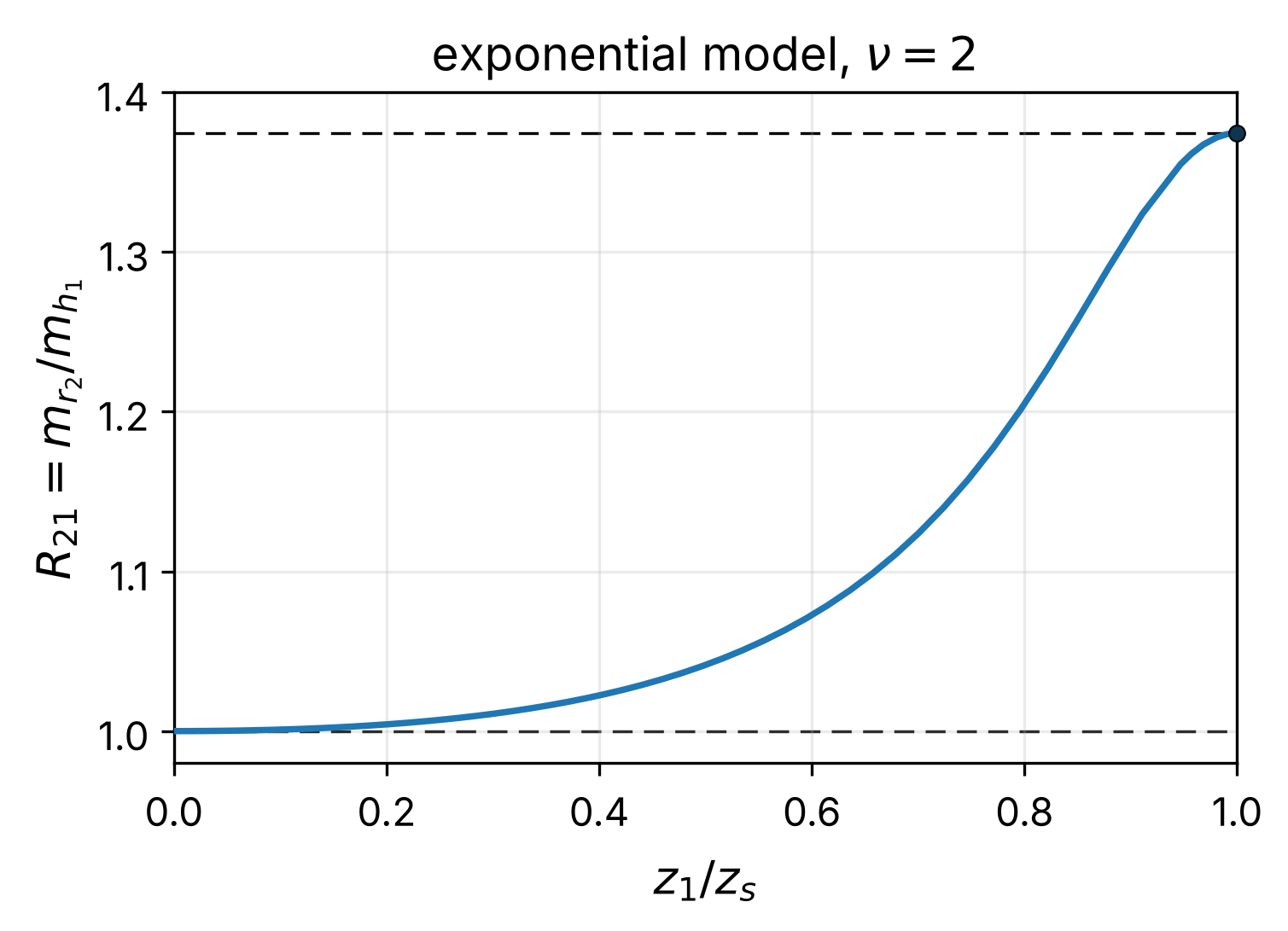}
\includegraphics[width=0.49\textwidth]{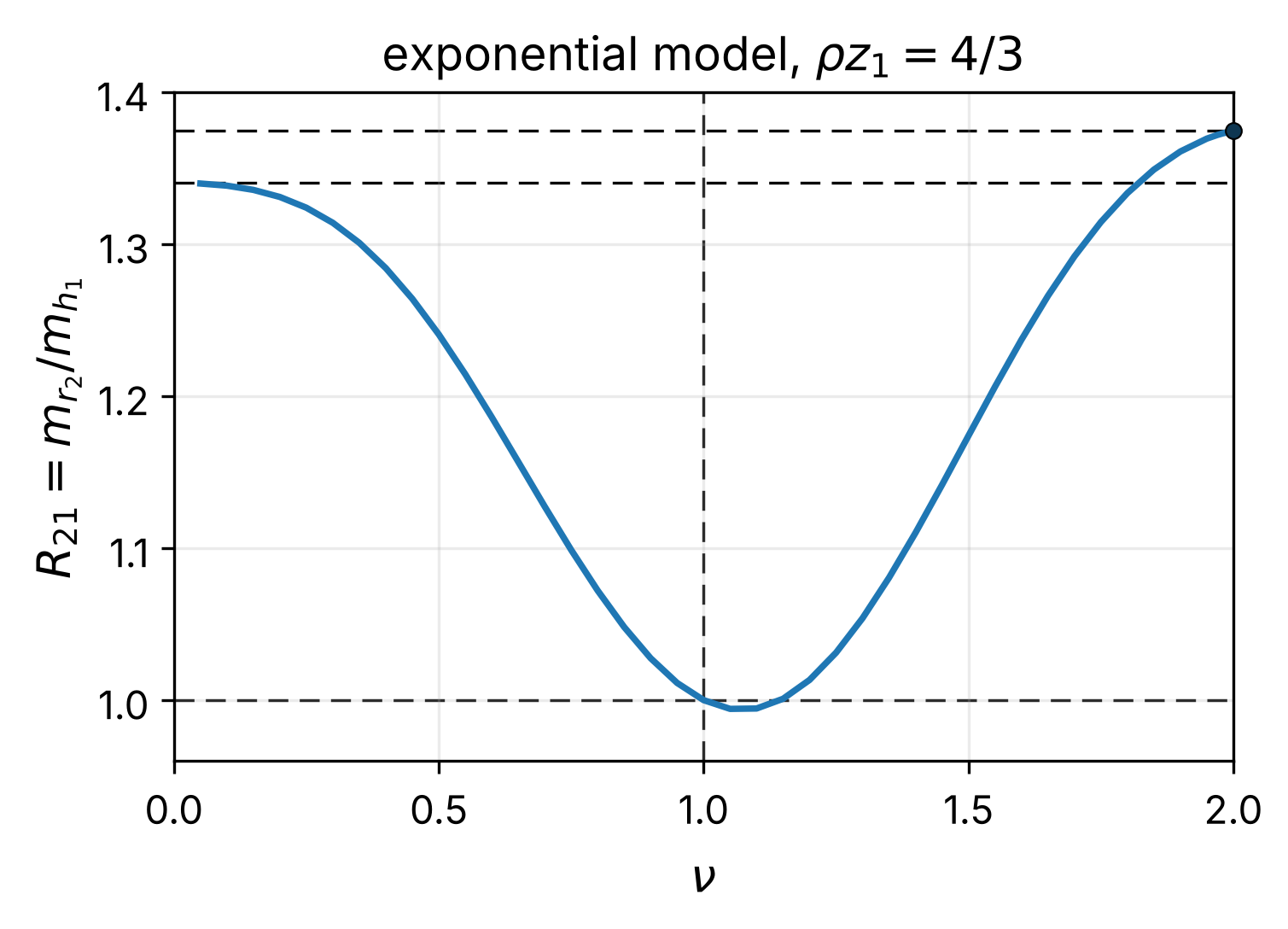}
\end{center}
\caption{\it $R_{21}=m_{r_2}/m_{h_1}$ in the exponential model. Left: as a function of $z_1/z_s$ for
$\nu=2$. Right: as a function of $\nu$ for $\rho z_1=\rho z_s(\nu=2)=4/3$. The dot marks the
maximum of $R_{21}$, reached for $z_1\to z_s$ at $\nu=2$, and the upper horizontal dashed line its
value. In the right panel the lower dashed line above $R_{21}=1$ marks the value for $\nu\to0$,
and the vertical dashed line marks $\nu=1$.}
\label{R21_max}
\end{figure}

It is interesting to find the maximum value of $R_{21}$. This maximum, $R_{21}\simeq1.37$, is
reached at $\nu=2$ when the IR brane approaches the singularity, $z_1\to z_s$ (see App.~\ref{Ap:unitarity}). In this limit the
masses can be obtained analytically, in terms of zeros of Bessel functions.
Figure~\ref{R21_max} shows this maximum in two ways. On the left panel, $R_{21}$ is plotted as a
function of $z_1/z_s$ for $\nu=2$: it increases monotonically from $R_{21}\to1$ for
$z_1/z_s\to0$ up to its maximum for $z_1/z_s\to1$ (dot). On the right panel, $R_{21}$ is plotted as a
function of $\nu$ at the fixed value $\rho z_1=\rho z_s(\nu=2)=4/3$, for which $\nu_{\max}=2$, so
the curve ends exactly at $\nu=2$. Starting from its value for $\nu\to0$ (lower horizontal dashed
line), $R_{21}$ decreases, takes the value $1$ at $\nu=1$, and then increases up to the maximum at
$\nu=2$ (upper horizontal dashed line), which is larger than the value for $\nu\to0$.

\section{Analytical derivation of the upper bound on $R_{11}$}
\label{sec:maxR11}
For the exponential superpotential model introduced in Sec.~\ref{subsec:exponential}, we now derive analytically the upper bound on $R_{11}$, the ratio between the lightest scalar mass and the lightest massive graviton mass. We restrict the discussion to Neumann-Neumann (N-N) BCs for the graviton. The argument combines two ingredients. First, the graviton and scalar Schr\"odinger operators of Eqs.~\eqref{eq:Sch_F} and~\eqref{eq:Sch_h} can be factorized in terms of first-order differential operators, which maps the positive graviton spectrum to a Dirichlet problem for the scalar bulk operator. Second, the scalar Schr\"odinger wave function $\widetilde{F}$ obeys Robin rather than Dirichlet (D) BCs. The two scalar problems therefore have the same bulk differential operator but different admissible function spaces, and their lowest eigenvalues can be compared by the Rayleigh--Ritz variational principle.
The first part of the argument proves $R_{11}<1$ at every regular point. We then use the endpoint behavior derived in App.~\ref{Ap:KK} to show that regular values of $R_{11}$ approach 1 arbitrarily closely. Hence $1$ is the sharp upper bound, $\sup R_{11}=1$.

\subsection{Variational setting for the exponential superpotential model}
We work on the regular finite interval $I=[0,z_1]$, with $0<z_1<z_s$, and approach the singular endpoint only after establishing the variational inequality. The Hilbert space is $\mathcal H=L^2(I)$, with inner product
\begin{equation}
 \langle\phi,\psi\rangle_{\mathcal H}
 =\int_0^{z_1}\phi^*(z)\psi(z)\,dz.
\end{equation}

To formulate the variational comparison, consider first a regular Schr\"odinger operator
\begin{equation}
 H=-\frac{d^2}{dz^2}+V(z)
\end{equation}
with Robin BCs
\begin{equation}
 \psi'(0)+b_0\psi(0)=0,
 \qquad
 \psi'(z_1)+b_1\psi(z_1)=0.
\end{equation}
The quantity minimized in the Rayleigh-Ritz principle is the quadratic form associated with this problem~\cite{Kato:1995},
\begin{equation}
 q[\psi]
 =\int_0^{z_1}\left(|\psi'|^2+V|\psi|^2\right)dz
 +b_1|\psi(z_1)|^2-b_0|\psi(0)|^2.
\end{equation}

We denote by $\mathcal D(\cdot)$ the domain of an operator or quadratic form. For the quadratic form $q$ defined above, the corresponding domain is the Sobolev space $\mathcal D(q)=H^1(I)=\{\psi\in\mathcal H:\psi'\in\mathcal H\}$. In one dimension these functions have well-defined endpoint values, so the boundary terms are finite.
For later use, the Dirichlet subspace is
\begin{equation}
 H_0^1(I)=\left\{\psi\in H^1(I):\psi(0)=\psi(z_1)=0\right\},
 \qquad H_0^1(I)\subset H^1(I).
\end{equation}

For a regular self-adjoint Sturm-Liouville problem, the lowest eigenvalue is then characterized by the Rayleigh--Ritz variational principle~\cite{Teschl:2014},
\begin{equation}
 \lambda_1
 =\inf_{\psi\in\mathcal D(q)\setminus\{0\}}
 \frac{q[\psi]}{\|\psi\|_{\mathcal H}^2}.
\end{equation}

In the Schr\"odinger problems considered here the eigenvalue is the squared mass, $\lambda=m^2$. The bound below follows from applying this statement to two realizations of the same scalar bulk operator with different domains.
\subsection{Factorization of the graviton and scalar operators}
Using the graviton and scalar Schrödinger operators in Eqs.~(\ref{eq:Sch_F}) and~(\ref{eq:Sch_h}), we define the first-order operator
\begin{equation}
 Q_z
 =\frac{d}{dz}
 +\frac{3\rho}{2\nu^2(1-\bar\rho z)},
 \qquad
 \mathcal D(Q_z)=H^1(I).
\end{equation}

The coefficient multiplying the wave function is bounded on every regular interval $0<z_1<z_s$, so $Q_z$ maps $H^1(I)$ into $\mathcal H$.
The adjoint must be determined together with its domain. For $\phi,\psi\in H^1(I)$, integration by parts gives
\begin{align}
 \langle\phi,Q_z\psi\rangle_{\mathcal H}
 &=\left[\phi^*(z)\psi(z)\right]_0^{z_1}
 +\int_0^{z_1}
 \left[
 -\phi'(z)
 +\frac{3\rho}{2\nu^2(1-\bar\rho z)}\phi(z)
 \right]^*\psi(z)\,dz.
\end{align}
Because $\mathcal D(Q_z)=H^1(I)$ imposes no BCs on $\psi$, the boundary term must vanish for arbitrary endpoint values, which requires $\phi(0)=\phi(z_1)=0$. Therefore
\begin{equation}
 Q_z^\dagger
 =-\frac{d}{dz}
 +\frac{3\rho}{2\nu^2(1-\bar\rho z)},
 \qquad
 \mathcal D(Q_z^\dagger)=H_0^1(I).
\end{equation}
A direct multiplication gives
\begin{equation}
H_h=Q_z^\dagger Q_z,
\qquad
H_F=Q_z Q_z^\dagger.
\label{eq:factorization}
\end{equation}
Thus the graviton and scalar bulk operators are related by an exact factorization on every regular finite interval.

\subsection{Massive N-N graviton spectrum and the D-D scalar problem}

For the N-N problem, the graviton BCs can be written as
\begin{equation}
 Q_z \widetilde{h}=0
 \qquad \text{at }z=0,z_1.
\end{equation}

Let $\widetilde{h}_n$ be a massive graviton eigenfunction satisfying $H_h\widetilde{h}_n=m_{h_n}^2\widetilde{h}_n$ with $m_{h_n}^2>0$, and define $\hat{F}_n=Q_z\widetilde{h}_n$. Using $H_h=Q_z^\dagger Q_z$ and $H_F=Q_zQ_z^\dagger$,
\begin{equation}
 H_F\hat{F}_n=Q_zH_h\widetilde{h}_n=m_{h_n}^2\hat{F}_n,
 \qquad
 \hat{F}_n(0)=\hat{F}_n(z_1)=0,
\end{equation}
so every positive graviton eigenvalue belongs to the D-D spectrum of $H_F$. Conversely, acting with $Q_z^\dagger$ on any positive D-D eigenfunction of $H_F$ gives the reverse map. If $H_F\hat{F}_n=m_{h_n}^2\hat{F}_n$ with $\hat{F}_n(0)=\hat{F}_n(z_1)=0$ and $m_{h_n}^2>0$, then $\widetilde{h}_n=Q_z^\dagger \hat{F}_n$ satisfies $H_h\widetilde{h}_n=m_{h_n}^2\widetilde{h}_n$, while $Q_z\widetilde{h}_n=m_{h_n}^2\hat{F}_n$ enforces the N-N BCs at both endpoints. Therefore the positive graviton spectrum coincides with the D-D spectrum of $H_F$; the zero mode is exceptional because $Q_z\widetilde{h}_0=0$. In particular, the lightest massive graviton is the lowest D-D eigenvalue of $H_F$, so
\begin{equation}
 m_{h_1}^2
 =\inf_{\hat{F}\in H_0^1(I)\setminus\{0\}}
 \frac{q_{DD}[\hat{F}]}{\|\hat{F}\|_{\mathcal H}^2},
\end{equation}
where
\begin{equation}
 q_{DD}[\hat{F}]
 =\int_0^{z_1}\left(|\hat{F}'|^2+V_F|\hat{F}|^2\right)dz,
 \qquad
 \mathcal D(q_{DD})=H_0^1(I).
\end{equation}

\subsection{Variational comparison with the physical scalar problem}
The scalar Schrödinger wave function $\widetilde{F}$ obeys the Robin boundary conditions given in Eq.~(\ref{eq:BC_radion_graviton}). The corresponding quadratic form is
\begin{align}
 q_R[\widetilde{F}]
 &=\int_0^{z_1}\left(|\widetilde{F}'|^2+V_F|\widetilde{F}|^2\right)dz
 +\frac{(2\nu^2-3)\rho}{2\nu^2(1-\bar\rho z_1)}|\widetilde{F}(z_1)|^2
 -\frac{(2\nu^2-3)\rho}{2\nu^2}|\widetilde{F}(0)|^2,
\end{align}
with $\mathcal D(q_R)=H^1(I)$. Since $H_0^1(I)\subset H^1(I)$ and the endpoint values vanish for every $\widetilde{F}\in H_0^1(I)$, one has $q_R[\widetilde{F}]=q_{DD}[\widetilde{F}]$ on the Dirichlet subspace. Thus the D-D variational problem is the restriction of the physical Robin problem to the smaller domain $H_0^1(I)$. If $m_{r_1}$ denotes the lightest physical scalar mass, Rayleigh--Ritz gives
\begin{align}
 m_{r_1}^2
 &=\inf_{\widetilde{F}\in H^1(I)\setminus\{0\}}
 \frac{q_R[\widetilde{F}]}{\|\widetilde{F}\|_{\mathcal H}^2}\notag  \\
 &\leq
 \inf_{\widetilde{F}\in H_0^1(I)\setminus\{0\}}
 \frac{q_R[\widetilde{F}]}{\|\widetilde{F}\|_{\mathcal H}^2}\notag  \\
 &=\inf_{\widetilde{F}\in H_0^1(I)\setminus\{0\}}
 \frac{q_{DD}[\widetilde{F}]}{\|\widetilde{F}\|_{\mathcal H}^2}
 =m_{h_1}^2 \,.
\end{align}
Consequently,
\begin{equation}
 R_{11}\equiv\frac{m_{r_1}}{m_{h_1}}\leq1 \,.
\end{equation}
The inequality is strict on every regular finite interval. If equality held, the normalized D-D ground state would also minimize the physical Robin functional and would satisfy $\widetilde{F}(0)=\widetilde{F}'(0)=0$. Uniqueness of the regular second-order initial-value problem would then force $\widetilde{F}\equiv0$, contradicting normalization. Hence
\begin{equation}
 R_{11}<1
\end{equation}
throughout the regular parameter domain.

\subsection{Sharpness of the upper bound}
The variational argument above shows that $R_{11}<1$ for every regular configuration with a finite IR brane position. To determine whether this upper bound can be improved, we now approach the singular endpoint. Since $\bar\rho=(\nu^2-1)\rho/\nu^2$, its conformal position $z_s=1/\bar\rho\equiv z_s(\nu)$ depends on $\nu$. The endpoint analysis of App.~\ref{Ap:KK} gives
\begin{equation}
\lim_{\nu\to1^+}
\left[
\lim_{z_1\to z_s(\nu)^-}R_{11}(\nu,z_1)
\right]
=1 \,.
\label{eq:R11_endpoint_limit}
\end{equation}
Hence there are regular configurations with $z_1<z_s(\nu)$ for which $R_{11}$ is arbitrarily close to one. Since $R_{11}<1$ at every regular point, the bound is sharp:
\begin{equation}
\sup R_{11}=1 \,. \label{eq:R11_sup_exp}
\end{equation}
The value $R_{11}=1$ is not attained at any regular finite configuration, only approached in the singular-endpoint limit. The result above concerns the exponential superpotential model. For the polynomial superpotential, the numerical analysis shows that the maximal value of $R_{11}$ increases with $n$. In the limit $n\to\infty$, the polynomial model reproduces the exponential superpotential model and therefore reaches the same limiting upper bound,
\begin{equation}
\sup_{n} R_{11}=1 \,. \label{eq:R11_sup_polynomial}
\end{equation}

\section{Discussion on unitarity constraints}
\label{sec:unitarity}
The results obtained for the exponential model show two different behaviors for the first two scalar mass ratios. The upper bound on $R_{11}$ is reached in the limit discussed in Sec.~\ref{sec:maxR11}, while $R_{21}$ reaches a larger maximum at $\nu=2$ when the IR brane approaches the singular endpoint, as shown in App.~\ref{Ap:unitarity}. This motivates the analysis of these results in relation to the unitarity sum rule of Ref.~\cite{Bertucci:2026osd}, by which a necessary condition for unitarity of the scattering amplitude $\mathcal M(h_1h_1\to h_1h_1)$, in the presence of $N$ scalar modes, is given by
\be
S_N=\sum_{\ell=1}^N \left(R_{\ell 1}^2-\frac{4}{3} \right)\left(\frac{c_\ell}{c_1}\right)^2<0\,.
\label{eq:unitarity_constraint}
\ee

In this sum rule, the scalar contribution depends on the mass ratios $R_{\ell 1}$ and on the trilinear couplings $c_\ell$ of the scalar modes to two first massive gravitons. For the first scalar mode, the upper bound obtained in Sec.~\ref{sec:maxR11} gives $R_{11}<1$, and therefore its contribution enters with a negative factor $R_{11}^2-4/3<-1/3$. 

For the second scalar mode, the situation is different. In the exponential model, $R_{21}$ reaches its maximum at $\nu=2$ and $z_1\to z_s$, as we can see from Fig.~\ref{R21_max}. As shown in Eq.~(\ref{eq:appendixB_R21_max}), $R_{21}^{\rm max}=1.3744$, and therefore $\left(R_{21}^{\rm max}\right)^2>4/3$. The contribution of this mode therefore has to be considered together with its trilinear coupling and with the contribution of the first scalar mode. App.~\ref{Ap:unitarity} shows that the first two scalar modes already give $S_2\simeq -0.68735<0$, as shown in Eq.~(\ref{eq:appendixB_unitarity_S2}).

The contribution of the higher scalar modes is included through the quantity $S_N$ defined in Eq.~(\ref{eq:unitarity_constraint}). The numerical sum converges to
\[
S_\infty\simeq -0.2966<0 \, ,
\]
as shown in Eq.~(\ref{eq:appendixB_unitarity_Sinf}). Thus, for the exponential model at $\nu=2$ and $z_1\to z_s$, the scalar contribution remains negative when the complete scalar KK tower is included, as required by the unitarity sum rule.

The analysis in App.~\ref{Ap:unitarity} is restricted to the configuration of the exponential model. The extension to generic values of $\nu$ and $z_1$, as well as to the power like model, is left for future work.

\section{Conclusions and outlook}
\label{sec:conclusions}
    
     In this paper we have computed the upper bound on the mass ratio between the lightest radion mode and the first graviton mode, $R_{11}$, in five dimensional models with a warped extra dimension, which we have associated with the rigidity of the internal manifold: the orbifold $S^1/\mathbb Z_2$. We have analyzed models with a consistent naked singularity in the metric and found a correlation between the upper bound on $R_{11}$ and the strength of the singularity of the Ricci scalar. We have analyzed the models with a singular behavior, in particular those with power law superpotential, $W\sim\bar\phi^n$ (with $n>2$), and those with exponential superpotential $W\sim e^{\nu\bar\phi}$ (with $\nu\geq 0$). We have found that the maximum value of $R_{11}(n)$ increases with increasing values of $n$, according with the strength of the Ricci scalar singularity $R\sim (\log z)^n$, while in the limit $n\to \infty$, since the background solution goes to the solution of the model with exponential superpotential, $R_{11}(n)$ goes to the value of $R_{11}(\nu)$, and the Ricci scalar tends to that of the exponential model. On the other hand, we have found in the exponential model that the maximum value of $R_{11}(\nu)$ increases (decreases) with $\nu$ for $\nu<1$ ($\nu>1$) and has its maximum value for $\nu\to1^{+}$, in agreement with the behavior of the Ricci scalar $R\sim z^{2\nu^2/(|\nu^2-1|)}$. Moreover, $R_{11}(\nu)$ gets its maximum value for $\nu\to1^+$, approaching the metric known as the linear dilaton background, where the spectrum of bulk fields is a gapped continuum. This behavior is consistent with the fact that the singularity strength of the Ricci scalar increases exponentially $R\sim e^{2\rho z}$, as compared with the power law increase of models with $\nu\neq 1$, or even with the logarithmic increase of models with power law superpotentials.  Power law models with $n\leq 2$ provide smaller values of $R_{11}$, while exponential models with $\nu\geq 2$ exhibit a bad naked singularity, so all possible models have been considered in our study. Moreover, we have also provided bounds for the ratio of the next to lightest radion mode and the first graviton mode, $R_{21}$. 
    
    The final results are summarized in Eqs.~(\ref{eq:R11}) and (\ref{eq:R21}). Note that the inequalities are saturated when the IR brane $z_1$ and the singularity $z_s$ coincide. However the singularity is clearly outside the range of validity of the effective field theory. In fact, higher curvature terms in the 5D action are suppressed by powers of $R/\Lambda_5^2$, where the cutoff $\Lambda_5$ is the energy scale at which the 5D gravity theory becomes strongly coupled. A Naive Dimensional Analysis (NDA) estimate provides $\Lambda_5^3\sim 24 \pi^3 M_5^3 $~\cite{Agashe:2007zd}. This means that inequalities in Eqs.~(\ref{eq:R11}) and (\ref{eq:R21}) are strict. 
    
    As the maximum value of $R_{11}$ takes place for the exponential superpotential corresponding to $\nu\to 1^+$, which also corresponds to the Ricci scalar with the strongest divergence, we can quantify an upper bound on $z_1$ for the linear dilaton model, which will also be a conservative upper bound for all considered models.  Using the value of the Ricci scalar, the condition on the location of the IR brane
    \be
    \left| R(z_1,\nu=1)\right|=12 \rho^2e^{2\rho z_1}<\Lambda_5^2
    \ee
    translates into the upper bound on $z_1$ given by
    \be
    \rho z_1<0.96+\frac{2}{3}\log\frac{M_4}{\rho} \,,
    \label{eq:bound_z1}
    \ee
   where we have used the relation between the 5D and 4D Planck scales, $M^3_5=\rho M_4^2$ (see e.g.~Ref.~\cite{Megias:2026wwp}). In particular for $\rho\simeq 1$ TeV, the bound (\ref{eq:bound_z1}) translates into the upper limit $\rho z_1\lesssim 36.4$, which is widely satisfied in all our numerical analysis. 
   
   Finally as the radion and all its modes have a trilinear coupling to two gravitons $h_n h_n$, they will contribute to the unitarization of the scattering amplitude $\mathcal M(h_n h_n\to h_n h_n)$ which must behave for high energies as $\mathcal M\sim s$. This point has been recently emphasized for the effective theory with a tower of scalars coupled to the graviton in Ref.~\cite{Bertucci:2026osd}, where unitarization of the scattering amplitude $\mathcal M(h_1 h_1\to h_1 h_1)$ has been imposed. 
   We find the upper limit on the mass $m_{r_1}$ satisfies the unitarization conditions. However including in the effective theory higher radion modes $r_n$, the role of the trilinear couplings has to be taken into account for imposing perturbative unitarity. This point that has been shown for the particular case of $\nu=2$ and $z_1=z_s$ in the exponential superpotential model, will be explored in more generality in the future.  

   \begin{acknowledgments}
LS thanks Alex Pomarol for discussions that motivated this work and IFAE for hospitality during the initial stages of this work.
EM would like to thank the Instituto de F\'{\i}sica of the Universidade de S\~ao Paulo, Brazil, and the ICTP South American Institute for Fundamental Research (SAIFR), S\~ao Paulo, Brazil, for their hospitality and financial support during the initial stages of this work.
The work of LD has been partially supported by CONICET, grant PIP-11220200101426.
LS is supported by CONICET and CLAF.
%
The work of EM and MQ is supported by the ``Proyectos de Investigaci\'on Precompetitivos'' Program of the Plan Propio de Investigaci\'on of the University of Granada.
The research of MQ is also supported by the grant PID2023-146686NB-C31 funded by
MICIU/AEI/10.13039/501100011033/ and by ERDF/EU, and under Severo Ochoa Centres of Excellence Programme 2025-2029 (CEX2024001442-S). 
IFAE is partially funded by the CERCA program of the Generalitat de Catalunya.

\end{acknowledgments}

\appendix

\section{KK modes and mass ratio limits in the exponential model}
\label{Ap:KK}

This appendix presents the KK analysis of the exponential model. In Sec.~\ref{app:bessel_exp}, for $\nu\neq1$, we solve the scalar and graviton Schr\"odinger equations in terms of Bessel functions and derive the corresponding mass conditions. In Sec.~\ref{app:mass_ratio_limits}, we use these results to study the singular endpoint limits on the $\nu<1$ and $\nu>1$ branches and derive the limiting mass ratios as $\nu\to1^\pm$. Finally, in Sec.~\ref{app:linear_dilaton}, we consider separately the linear dilaton model, $\nu=1$, and determine the scalar and graviton spectra and the corresponding mass ratios.

\subsection{Bessel solutions and boundary conditions for $\nu\neq1$}\label{app:bessel_exp}
For $\nu\neq1$, substituting the exponential model potentials in Eqs.~(\ref{eq:Vh_exp}) and~(\ref{eq:VF_exp}) into the Schr\"odinger equations~(\ref{eq:Sch_F}) and~(\ref{eq:Sch_h}) gives as solutions linear combinations of Bessel functions. Their general solutions can be written as
\begin{align}
\widetilde{h}(z)
&=
\sqrt{1-\bar\rho z}
\left[
 a_h J_{\beta_h}\!\left(d_h(z)\right)
 +
 b_h Y_{\beta_h}\!\left(d_h(z)\right)
\right],\\
\widetilde{F}(z)
&=
\sqrt{1-\bar\rho z}
\left[
 a_F J_{\beta_F}\!\left(d_F(z)\right)
 +
 b_F Y_{\beta_F}\!\left(d_F(z)\right)
\right].
\end{align}
The Bessel arguments are
\begin{equation}
d_h(z)=\frac{m_h}{|\bar\rho|}\left(1-\bar\rho z\right) \,,
\qquad
 d_F(z)=\frac{m_r}{|\bar\rho|}\left(1-\bar\rho z\right) \,,
\label{eq:bessel_arguments_sec3}
\end{equation}
and the corresponding orders are
\begin{equation}
\beta_h=\frac{4-\nu^2}{2|\nu^2-1|} \,,
\qquad
\beta_F=\frac{\nu^2+2}{2|\nu^2-1|} \,.
\label{eq:bessel_orders_sec3}
\end{equation}

The brane boundary conditions in Eq.~(\ref{eq:BC_radion_graviton}), expressed in terms of $\tilde F$ and $\tilde h$ through Eqs.~(\ref{eq:Sch_F}) and~(\ref{eq:Sch_h}), lead to the following boundary operators for the Bessel solutions:
\begin{align}
B_h[Z](d)
&=
 d Z_{\beta_h}'(d)
 +
 \left[
 \frac{1}{2}
 -
 \frac{3}{2(\nu^2-1)}
 \right]
 Z_{\beta_h}(d) \,, 
\label{eq:Bh_sec3}\\
B_F[Z](d)
&=
 d Z_{\beta_F}'(d)
 +
 \frac{2-\nu^2}{2(\nu^2-1)}
 Z_{\beta_F}(d) \,,
\qquad Z=J,Y \,.
\label{eq:BF_sec3}
\end{align}
The boundary conditions are then simply
\begin{equation}
B_h[\widetilde{h}]=0,
\qquad
B_F[\widetilde{F}]=0,
\qquad
z=0,z_1.
\label{eq:BC_B_sec3}
\end{equation}
Using the standard Bessel recurrence relations~\cite{Watson:1944}
\begin{equation}
 dZ_\beta'(d)+\beta Z_\beta(d)=dZ_{\beta-1}(d),
\qquad
 dZ_\beta'(d)-\beta Z_\beta(d)=-dZ_{\beta+1}(d),
\label{eq:bessel_rec_sec3}
\end{equation}
the boundary operators simplify on the two branches.
\begin{itemize}
\item
For $\nu<1$, one has $\beta_h=\beta_F+1$, and
\begin{equation}
B_h[Z](d)=dZ_{\beta_F}(d),
\qquad
B_F[Z](d)
=
 dZ_{\beta_F-1}(d)
 -
 \frac{2}{1-\nu^2}Z_{\beta_F}(d).
\label{eq:branch_left_sec3}
\end{equation}
\item
For $\nu>1$, one has $\beta_h=\beta_F-1$, and
\begin{equation}
B_h[Z](d)=-dZ_{\beta_F}(d),
\qquad
B_F[Z](d)
=
 -dZ_{\beta_F+1}(d)
 +
 \frac{2}{\nu^2-1}Z_{\beta_F}(d).
\label{eq:branch_right_sec3}
\end{equation}
\end{itemize}

Imposing the boundary conditions at both endpoints, a nontrivial solution exists only if the corresponding mass condition is satisfied. For the N-N graviton spectrum,
\begin{equation}
J_{\beta_F}\!\left(d_h(0)\right)
Y_{\beta_F}\!\left(d_h(z_1)\right)
-
Y_{\beta_F}\!\left(d_h(0)\right)
J_{\beta_F}\!\left(d_h(z_1)\right)
=0\,
\label{eq:graviton_spectral_sec3}
\end{equation}
while the scalar spectrum is determined by
\begin{equation}
B_F[J]\!\left(d_F(0)\right)
B_F[Y]\!\left(d_F(z_1)\right)
-
B_F[Y]\!\left(d_F(0)\right)
B_F[J]\!\left(d_F(z_1)\right)
=0.
\label{eq:scalar_spectral_sec3}
\end{equation}

\subsection{Mass ratio limits as $\nu\to1^\pm$}\label{app:mass_ratio_limits}

We now derive the limiting mass ratios as $\nu\to1^\pm$. For $\nu<1$, the singular endpoint lies at $z=\infty$, while for $\nu>1$ it lies at the finite conformal position $z_s=1/\bar\rho$. The two limits to be established are
\begin{equation}
\lim_{\nu\to1^-}
\left[
\lim_{z_1\to\infty}R_{11}(\nu,z_1)
\right]
=
\frac{2\sqrt{2}}{3},
\qquad
\lim_{\nu\to1^+}
\left[
\lim_{z_1\to z_s(\nu)^-}R_{11}(\nu,z_1)
\right]
=1.
\label{eq:two_endpoint_limits}
\end{equation}

For these limits, we first take the singular endpoint limit at fixed $\nu$ and only then let $\nu\to1^\pm$. We use the Bessel representation introduced in Sec.~\ref{app:bessel_exp}. In particular, the graviton and scalar Bessel arguments are those defined in Eq.~\eqref{eq:bessel_arguments_sec3}. Their behavior at the two endpoints differs on the two sides of $\nu=1$, leading to distinct limiting mass conditions.

\subsubsection{The $\nu<1$ branch}

\paragraph{Graviton spectrum.}
For $\nu<1$, one has $\bar\rho<0$, and the singular endpoint is reached as $z_1\to\infty$. The Bessel arguments are those of Eq.~\eqref{eq:bessel_arguments_sec3}; therefore their values at the two endpoints satisfy the exact relation
\begin{equation}
d_h(0)=
\frac{d_h(z_1)}{1-\bar\rho z_1}.
\label{eq:dG_endpoint_relation}
\end{equation}
For each graviton level $n$, we assume that $d_{h_n}(z_1)$ remains finite in the endpoint limit. The result below will show that this assumption is consistent. Since $1-\bar\rho z_1\to\infty$, Eq.~\eqref{eq:dG_endpoint_relation} implies $d_{h_n}(0)\to0$. For positive Bessel order $\beta$, the small argument behavior is~\cite{Watson:1944}
\begin{equation}
J_\beta(w)
\simeq
\frac{(w/2)^\beta}{\Gamma(\beta+1)},
\qquad
Y_\beta(w)
\simeq
-\frac{\Gamma(\beta)}{\pi}
\left(\frac{2}{w}\right)^\beta,
\qquad
w\to0^+.
\label{eq:bessel_small_w}
\end{equation}
Hence $J_{\beta_F}(d_{h_n}(0))/Y_{\beta_F}(d_{h_n}(0))\to0$. The N-N graviton mass condition is Eq.~\eqref{eq:graviton_spectral_sec3}. For the finite endpoint roots considered here, $Y_{\beta_F}(d_{h_n}(z_1))$ remains finite. Dividing Eq.~\eqref{eq:graviton_spectral_sec3} by $Y_{\beta_F}(d_{h_n}(0))$ and taking the endpoint limit gives $J_{\beta_F}(d_{h_n}(z_1))\to0$. Thus $d_{h_n}$ are the zeros of the Bessel function $J_{\beta_F}$, and
\begin{equation}
\lim_{z_1\to\infty}d_{h_n}(z_1)
=
j_{\beta_F,n}.
\label{eq:dG_limit_left}
\end{equation}
Equation~\eqref{eq:dG_limit_left} confirms the consistency of the assumption that $d_{h_n}(z_1)$ remains finite.

Having taken the endpoint limit, we now consider $\nu\to1^-$. Using the definition of $\beta_F$ in Eq.~\eqref{eq:bessel_orders_sec3},
\begin{equation}
\beta_F\simeq\frac{3}{2(1-\nu^2)}.
\label{eq:beta_large_left}
\end{equation}
For the lightest massive graviton, the first positive Bessel zero has the large order expansion~\cite{Olver:1951}
\begin{equation}
j_{\beta_F,1}
=
\beta_F+a_1\beta_F^{1/3}+\cdots,
\label{eq:first_bessel_zero_asymptotic}
\end{equation}
and therefore
\begin{equation}
\lim_{z_1\to\infty}d_{h_1}(z_1)
\simeq
\frac{3}{2(1-\nu^2)}
\qquad
(\nu\to1^-).
\label{eq:dG1_asymptotic_left}
\end{equation}

\paragraph{Scalar spectrum.}
For the scalar, the boundary condition is Eq.~\eqref{eq:BC_B_sec3}, with $B_F$ given by Eq.~\eqref{eq:branch_left_sec3}, and the masses follow from Eq.~\eqref{eq:scalar_spectral_sec3}. We again compare the Bessel arguments at $z=0$ and $z=z_1$. For each scalar level $\ell$, we assume that $d_{F_\ell}(z_1)$ remains finite in the endpoint limit. Equation~\eqref{eq:bessel_arguments_sec3} gives $d_F(0)=d_F(z_1)/(1-\bar\rho z_1)$, so $1-\bar\rho z_1\to\infty$ implies $d_{F_\ell}(0)\to0$. Using Eq.~\eqref{eq:bessel_small_w}, the ratio of the two UV boundary operators behaves as
\begin{equation}
\frac{B_F[J]\!\left(d_{F_\ell}(0)\right)}
{B_F[Y]\!\left(d_{F_\ell}(0)\right)}
\simeq d_{F_\ell}(0)^{2\beta_F}
\longrightarrow0.
\label{eq:BS_ratio_zero}
\end{equation}

For the finite endpoint roots considered here, $B_F[Y](d_{F_\ell}(z_1))$ remains finite. Dividing Eq.~\eqref{eq:scalar_spectral_sec3} by $B_F[Y](d_{F_\ell}(0))$ and taking the endpoint limit gives $B_F[J](d_{F_\ell}(z_1))=0$. Using the $\nu<1$ form of $B_F$ in Eq.~\eqref{eq:branch_left_sec3}, this becomes
\begin{equation}
d_{F_\ell}(z_1)
J_{\beta_F-1}\!\left(d_{F_\ell}(z_1)\right)
-
\frac{2}{1-\nu^2}
J_{\beta_F}\!\left(d_{F_\ell}(z_1)\right)
=0.
\label{eq:scalar_limit_equation_left}
\end{equation}
Equation~\eqref{eq:scalar_limit_equation_left} determines finite roots, confirming the consistency of the scalar assumption as well.
We now take $\nu\to1^-$, for which $\beta_F\to\infty$. The lightest scalar root lies in the region below the turning point~\footnote{For a Bessel equation of large order $\beta$, the turning point region is centered at $d\simeq\beta$. We refer to $d<\beta$ as the region below the turning point and to roots with $d/\beta\to1$ as belonging to the turning point region; see Ref.~\cite{Olver:1951}.}, where the large order relation gives, at leading order~\cite{Olver:1951},
\begin{equation}
d\,
\frac{J_{\beta-1}(d)}{J_\beta(d)}
\simeq
\beta+\sqrt{\beta^2-d^2}.
\label{eq:subturning_bessel_relation}
\end{equation}
Applying Eq.~\eqref{eq:subturning_bessel_relation} to Eq.~\eqref{eq:scalar_limit_equation_left} for the lightest scalar gives, at leading order,
\begin{equation}
\beta_F
+
\sqrt{\beta_F^2-d_{F_1}(z_1)^2}
\simeq
\frac{2}{1-\nu^2}.
\label{eq:scalar_subturning_equation}
\end{equation}
Using the leading behavior of $\beta_F$ in Eq.~\eqref{eq:beta_large_left}, one obtains directly
\begin{equation}
\lim_{z_1\to\infty}d_{F_1}(z_1)
\simeq
\frac{\sqrt{2}}{1-\nu^2}
\qquad
(\nu\to1^-).
\label{eq:dS1_asymptotic_left}
\end{equation}

At fixed $\nu$ and $z_1$, the physical masses and the endpoint Bessel arguments differ by the same multiplicative factor in the graviton and scalar sectors. This factor therefore cancels in $R_{11}$. Combining Eqs.~\eqref{eq:dG1_asymptotic_left} and \eqref{eq:dS1_asymptotic_left},
\begin{equation}
\lim_{\nu\to1^-}
\left[
\lim_{z_1\to\infty}R_{11}(\nu,z_1)
\right]
=
\lim_{\nu\to1^-}
\left[
\lim_{z_1\to\infty}
\frac{d_{F_1}(z_1)}{d_{h_1}(z_1)}
\right]
=
\frac{2\sqrt{2}}{3}.
\label{eq:R11_left_final}
\end{equation}

For every fixed scalar level $\ell\geq2$, the corresponding root belongs to the turning point tower and satisfies $d_{F_\ell}(z_1)\simeq\beta_F$ as $\nu\to1^-$. For every fixed graviton level $n$, Eq.~\eqref{eq:dG_limit_left} gives $d_{h_n}(z_1)=j_{\beta_F,n}\simeq\beta_F$ as $\nu\to1^-$. It follows that
\begin{equation}
\lim_{\nu\to1^-}
\left[
\lim_{z_1\to\infty}R_{\ell n}(\nu,z_1)
\right]
=1,
\qquad
\ell\geq2,
\quad n\geq1,
\label{eq:Relln_left_provisional}
\end{equation}
for fixed $\ell$ and $n$.

\subsubsection{The $\nu>1$ branch}

\paragraph{Graviton spectrum.}
For $\nu>1$, one has $\bar\rho>0$, and the singular endpoint lies at the finite conformal position $z_s$. The endpoint limit is therefore $z_1\to z_s^-$. In this branch the hierarchy of Bessel arguments is reversed. Using Eq.~\eqref{eq:bessel_arguments_sec3}, their endpoint values satisfy
\begin{equation}
d_h(z_1)=
\left(1-\bar\rho z_1\right)d_h(0).
\label{eq:dG_endpoint_relation_right}
\end{equation}
For each graviton level $n$, we assume that $d_{h_n}(0)$ remains finite in the endpoint limit. Since $1-\bar\rho z_1\to0$, Eq.~\eqref{eq:dG_endpoint_relation_right} implies $d_{h_n}(z_1)\to0$, and Eq.~\eqref{eq:bessel_small_w} gives $$J_{\beta_F}(d_{h_n}(z_1))/Y_{\beta_F}(d_{h_n}(z_1))\to0.$$ The N-N graviton mass condition is again Eq.~\eqref{eq:graviton_spectral_sec3}. For the finite endpoint roots considered here, $Y_{\beta_F}(d_{h_n}(0))$ remains finite. Dividing Eq.~\eqref{eq:graviton_spectral_sec3} by $Y_{\beta_F}(d_{h_n}(z_1))$ and taking the endpoint limit gives $J_{\beta_F}(d_{h_n}(0))\to0$. Hence
\begin{equation}
\lim_{z_1\to z_s(\nu)^-}d_{h_n}(0)
=
j_{\beta_F,n},
\label{eq:dG_limit_right}
\end{equation}
confirming the consistency of the assumption that $d_{h_n}(0)$ remains finite.

Having taken the endpoint limit, we now consider $\nu\to1^+$. From Eq.~\eqref{eq:bessel_orders_sec3}, $\beta_F\simeq3/[2(\nu^2-1)]$. For the lightest massive graviton, $j_{\beta_F,1}\simeq\beta_F$, and therefore
\begin{equation}
\lim_{z_1\to z_s(\nu)^-}d_{h_1}(0)
\simeq
\frac{3}{2(\nu^2-1)}
\qquad
(\nu\to1^+).
\label{eq:dG1_asymptotic_right}
\end{equation}

\paragraph{Scalar spectrum.}
For the scalar the hierarchy is likewise reversed. The boundary condition is Eq.~\eqref{eq:BC_B_sec3}, with $B_F$ given by Eq.~\eqref{eq:branch_right_sec3}, and the masses follow from Eq.~\eqref{eq:scalar_spectral_sec3}. For each fixed scalar level $\ell$, we assume that $d_{F_\ell}(0)$ remains finite in the endpoint limit. Equation~\eqref{eq:bessel_arguments_sec3} gives $d_F(z_1)=(1-\bar\rho z_1)d_F(0)$, which implies $d_{F_\ell}(z_1)\to0$. At small argument,
\begin{equation}
\frac{B_F[J]\!\left(d_{F_\ell}(z_1)\right)}
{B_F[Y]\!\left(d_{F_\ell}(z_1)\right)}
\simeq
d_{F_\ell}(z_1)^{2\beta_F}
\longrightarrow0.
\label{eq:BS_ratio_zero_right}
\end{equation}
For the finite endpoint roots considered here, $B_F[Y](d_{F_\ell}(0))$ remains finite. Dividing Eq.~\eqref{eq:scalar_spectral_sec3} by $B_F[Y](d_{F_\ell}(z_1))$ and taking the endpoint limit gives $B_F[J](d_{F_\ell}(0))=0$. Using the $\nu>1$ form of $B_F$ in Eq.~\eqref{eq:branch_right_sec3}, this is equivalently
\begin{equation}
d_{F_\ell}(0)
\frac{J_{\beta_F+1}\!\left(d_{F_\ell}(0)\right)}
{J_{\beta_F}\!\left(d_{F_\ell}(0)\right)}
=
\frac{2}{\nu^2-1}.
\label{eq:scalar_limit_equation_right}
\end{equation}
This equation determines finite roots, confirming the consistency of the scalar assumption.

We now take $\nu\to1^+$. Dividing the exact quantization condition \eqref{eq:scalar_limit_equation_right} by $\beta_F$ gives
\begin{equation}
\frac{d_{F_1}(0)}{\beta_F}
\frac{J_{\beta_F+1}\!\left(d_{F_1}(0)\right)}
{J_{\beta_F}\!\left(d_{F_1}(0)\right)}
=
\frac{4}{\nu^2+2}
\longrightarrow
\frac{4}{3}.
\label{eq:scalar_normalized_right}
\end{equation}
If the lightest scalar root remained parametrically below the turning point, the large order behavior would give
\begin{equation}
\frac{d}{\beta}
\frac{J_{\beta+1}(d)}{J_\beta(d)}
\simeq
1-\sqrt{1-\frac{d^2}{\beta^2}}
<1.
\label{eq:subturning_bessel_relation_right}
\end{equation}
This is incompatible with Eq.~\eqref{eq:scalar_normalized_right}, whose right-hand side tends to $4/3$. Thus $d_{F_1}(0)/\beta_F$ cannot approach a value below one.
The left side of Eq.~\eqref{eq:scalar_limit_equation_right} starts from zero at $d=0$ and diverges to $+\infty$ as $d$ approaches the first zero of $J_{\beta_F}$ from below. Hence the lightest root satisfies $d_{F_1}(0)<j_{\beta_F,1}$. For large $\beta_F$, $j_{\beta_F,1}=\beta_F+a_1\beta_F^{1/3}+\cdots$ with $a_1>0$, so that $j_{\beta_F,1}/\beta_F\to1$. Together with the exclusion of an asymptotic value below one, this gives $d_{F_1}(0)/\beta_F\to1$.
Consequently,
\begin{equation}
\lim_{z_1\to z_s(\nu)^-}d_{F_1}(0)
\simeq
\beta_F
\simeq
\frac{3}{2(\nu^2-1)}
\qquad
(\nu\to1^+).
\label{eq:dS1_asymptotic_right}
\end{equation}

The same multiplicative factor relates the physical masses to the endpoint Bessel arguments in both sectors and cancels in the ratio. Combining Eqs.~\eqref{eq:dG1_asymptotic_right} and \eqref{eq:dS1_asymptotic_right},
\begin{equation}
\lim_{\nu\to1^+}
\left[
\lim_{z_1\to z_s(\nu)^-}R_{11}(\nu,z_1)
\right]
=
\lim_{\nu\to1^+}
\left[
\lim_{z_1\to z_s(\nu)^-}
\frac{d_{F_1}(0)}{d_{h_1}(0)}
\right]
=1.
\label{eq:R11_right_final}
\end{equation}

The same reasoning applies to every fixed scalar level $\ell$. The $\ell$-th root is associated with the $\ell$-th pole of the scalar quantization condition, located at $j_{\beta_F,\ell}$, and for fixed $\ell$ one has $j_{\beta_F,\ell}\simeq\beta_F$ as $\nu\to1^+$. Hence all fixed scalar levels satisfy $d_{F_\ell}(0)\simeq\beta_F$. Likewise, for every fixed graviton level $n$, Eq.~\eqref{eq:dG_limit_right} gives $d_{h_n}(0)=j_{\beta_F,n}\simeq\beta_F$ as $\nu\to1^+$. Therefore,
\begin{equation}
\lim_{\nu\to1^+}
\left[
\lim_{z_1\to z_s(\nu)^-}R_{\ell n}(\nu,z_1)
\right]
=1,
\qquad
\ell\geq1,
\quad n\geq1,
\label{eq:Relln_right_provisional}
\end{equation}
for fixed $\ell$ and $n$.

\subsection{The $\nu=1$ linear dilaton model}\label{app:linear_dilaton}

For the case $\nu=1$, the background and Schr\"odinger potentials are given by Eqs.~(\ref{eq:linear_dilaton_background}) and~(\ref{eq:V_linear_dilaton}), respectively. Writing 
\be
m_g \equiv \frac{3}{2}\rho, 
\ee
Eqs.~(\ref{eq:Sch_F}) and~(\ref{eq:Sch_h}) reduce to constant coefficient equations. The scalar and graviton spectra are then fixed by imposing the corresponding boundary conditions in Eq.~(\ref{eq:BC_radion_graviton}).

\paragraph{Scalar spectrum.}
Using the scalar Schr\"odinger equation in Eq.~(\ref{eq:Sch_F}) together with the scalar boundary condition in Eq.~(\ref{eq:BC_radion_graviton}), the eigenvalue condition can be written as
\begin{equation}
\left(
 \sqrt{m_g^2-m_r^2}-\frac{m_g}{3}
\right) \times \left[
 1-e^{2\sqrt{m_g^2-m_r^2}\,z_1}
\right]
=0 \, .
\label{eq:IRBC}
\end{equation}
The first factor gives the lightest scalar mode,
\begin{equation}
 m_{r_1}=\frac{2\sqrt{2}}{3}\,m_g \, ,
 \label{eq:mrad_1}
\end{equation}
with wave function $F(z)\propto e^{2\rho z}$. The remaining modes follow from the second factor in Eq.~(\ref{eq:IRBC}),
\begin{equation}
 m_{r_\ell}^2
 =
 m_g^2+\frac{\pi^2(\ell-1)^2}{z_1^2} \, ,
 \qquad \ell=2,3,\ldots \, .
 \label{eq:mrad_n}
\end{equation}
The lightest mode has zero nodes, whereas the mode $r_\ell$ has $\ell-1$ nodes for $\ell\ge2$. The solution with $m_r=m_g$ is absent because its wave function is identically zero.

\paragraph{Graviton spectrum.}
For the graviton, Eq.~(\ref{eq:Sch_h}) together with the graviton boundary condition in Eq.~(\ref{eq:BC_radion_graviton}) gives, for the nonzero modes,
\begin{equation}
 1-e^{2\sqrt{m_g^2-m_h^2}\,z_1}=0 \, .
 \label{eq:graviton_linear_condition}
\end{equation}
There is also a massless graviton with a flat profile. The full spectrum is therefore
\begin{equation}
 m_h^{(0)}=0 \, ,
 \qquad
 m_{h_n}^2
 =
 m_g^2+\frac{\pi^2 n^2}{z_1^2} \, ,
 \qquad n=1,2,\ldots \, .
 \label{eq:spectrum_NN}
\end{equation}
As in the scalar sector, the solution with $m_h=m_g$ is absent because the corresponding wave function vanishes identically.

\paragraph{Mass ratios.}
Using Eqs.~(\ref{eq:mrad_1}) and~(\ref{eq:spectrum_NN}), one finds
\begin{equation}
 R_{1n}
 \equiv
 \frac{m_{r_1}}{m_{h_n}}
 =
 \frac{2\sqrt{2}}{3}
 \left[
 1 + \left( \frac{\pi n}{m_g z_1} \right)^2
 \right]^{-1/2} \, ,
 \qquad n\ge1 \, .
 \label{eq:R1n}
\end{equation}
For the higher scalar modes, Eqs.~(\ref{eq:mrad_n}) and~(\ref{eq:spectrum_NN}) give
\begin{equation}
 R_{\ell n}
 \equiv
 \frac{m_{r_\ell}}{m_{h_n}}
 =
 \left[
 \frac{
 1 + \left( \dfrac{\pi(\ell-1)}{m_gz_1} \right)^2
 }{
 1 + \left( \dfrac{\pi n }{m_g z_1 } \right)^2
 }
 \right]^{1/2} \, ,
 \qquad \ell\ge2 \, ,\quad n\ge1 \, .
 \label{eq:Rmn}
\end{equation}
It follows from Eqs.~(\ref{eq:R1n}) and~(\ref{eq:Rmn}) that
\begin{alignat}{3}
& R_{1n}
&& \xrightarrow[\rho z_1\to+\infty]{}
\frac{2\sqrt{2}}{3} \, ,
&\qquad& n\ge1 \, ,
\label{eq:R1n_linear_limit}
\\
& R_{\ell n}
&& \xrightarrow[\rho z_1\to+\infty]{}
1 \, ,
&\qquad& \ell\ge2,\quad n\ge1 \, ,
\label{eq:Rln_linear_limit}
\\
& R_{\ell,\ell-1}
&& =1 \, ,
&\qquad& \ell\ge2,\quad \text{for all } z_1 \, .
\label{eq:Rln_linear_exact}
\end{alignat}

\section{Maximum of $R_{21}$ and unitarity at the singular endpoint}
\label{Ap:unitarity}
In this appendix we study the exponential model at the singular endpoint. In the first part, we use the endpoint mass conditions derived in App.~\ref{Ap:KK} to determine $R_{21}$ and show that its maximum is reached for $\nu=2$ and $z_1\to z_s$. We then consider this limiting configuration in the context of the unitarity sum rule of Ref.~\cite{Bertucci:2026osd}. In particular, we evaluate the contribution of the scalar KK tower coupled to the first massive graviton and examine its compatibility with the scalar contribution required by the sum rule.
The mass conditions at the singular endpoint follow directly from the analysis of App.~\ref{Ap:KK}. 

For $\nu<1$, taking $z_1\to\infty$ gives, from Eq.~\eqref{eq:dG_limit_left},
\begin{equation}
 d_{h_n}(z_1)=j_{\beta_F,n},
 \qquad
 \beta_F=\frac{\nu^2+2}{2(1-\nu^2)}.
\label{eq:appendixB_graviton_left}
\end{equation}
The scalar masses are determined by Eq.~\eqref{eq:scalar_limit_equation_left}, namely
\begin{equation}
 d_{F_\ell}(z_1)
 J_{\beta_F-1}\!\left(d_{F_\ell}(z_1)\right)
 -\frac{2}{1-\nu^2}
 J_{\beta_F}\!\left(d_{F_\ell}(z_1)\right)=0.
\label{eq:appendixB_scalar_left}
\end{equation}
Therefore, on this branch,
\begin{equation}
 R_{21}(\nu)=
 \frac{d_{F_2}(z_1)}{j_{\beta_F,1}},
 \qquad \nu<1,
\label{eq:appendixB_R21_left}
\end{equation}
where $d_{F_2}(z_1)$ denotes the second physical scalar root of Eq.~\eqref{eq:appendixB_scalar_left}.

For $\nu>1$, taking $z_1\to z_s^-$ gives, from Eq.~\eqref{eq:dG_limit_right},
\begin{equation}
 d_{h_n}(0)=j_{\beta_F,n},
 \qquad
 \beta_F=\frac{\nu^2+2}{2(\nu^2-1)}.
\label{eq:appendixB_graviton_right}
\end{equation}
The scalar masses are determined by Eq.~\eqref{eq:scalar_limit_equation_right},
\begin{equation}
 d_{F_\ell}(0)
 \frac{J_{\beta_F+1}\!\left(d_{F_\ell}(0)\right)}
 {J_{\beta_F}\!\left(d_{F_\ell}(0)\right)}
 =\frac{2}{\nu^2-1}.
\label{eq:appendixB_scalar_right}
\end{equation}
Hence
\begin{equation}
 R_{21}(\nu)=
 \frac{d_{F_2}(0)}{j_{\beta_F,1}},
 \qquad \nu>1,
\label{eq:appendixB_R21_right}
\end{equation}
where $d_{F_2}(0)$ is the second scalar root of Eq.~\eqref{eq:appendixB_scalar_right}.

A numerical evaluation of these endpoint conditions in the range $0\leq\nu\leq2$ shows that $R_{21}$ reaches its maximum at $\nu=2$. At this point $\beta_F=1$, and the first massive graviton is fixed by
\begin{equation}
 d_{h_1}(0)=j_{1,1}=3.8317.
\label{eq:appendixB_nu2_graviton}
\end{equation}
The second scalar root is obtained from Eq.~\eqref{eq:appendixB_scalar_right}, which at $\nu=2$ becomes
\begin{equation}
 -d\,J_2(d)+\frac{2}{3}J_1(d)=0,
\label{eq:appendixB_nu2_scalar_eq}
\end{equation}
and gives
\begin{equation}
 d_{F_2}(0)=5.2663 .
\label{eq:appendixB_nu2_scalar}
\end{equation}
Consequently,
\begin{equation}
 R_{21}^{\rm max}=R_{21}(\nu=2)
 =\frac{5.2663}{3.8317}
 =1.3744.
\label{eq:appendixB_R21_max}
\end{equation}
\noindent
{\bf Unitarity.}
We consider the scalar contribution to the unitarity sum rule of Ref.~\cite{Bertucci:2026osd} for $\nu=2$ at the singular endpoint, where $R_{21}$ reaches the upper bound of the model given in Eq.~\eqref{eq:appendixB_R21_max}. The graviton and scalar masses are determined by Eqs.~\eqref{eq:appendixB_graviton_right} and \eqref{eq:appendixB_scalar_right}, respectively.
To evaluate this contribution, we need the couplings of the scalar modes to the lightest massive graviton. The corresponding profiles $h_1$ and $F_\ell$ are normalized according to
\begin{equation}
\int_0^{z_s} dz\,e^{-3A(z)}|\psi(z)|^2=1,
\qquad
\psi=F_\ell,h_1 \,,
\label{eq:appendixB_unitarity_norm}
\end{equation}
and are given by
\begin{equation}
h_1(z)=
\sqrt{\frac{3\rho}{2J_0^2(d_{h_1})}}\,
J_0\left[
d_{h_1}\left(1-\frac{3}{4}\rho z\right)
\right] \,,
\label{eq:appendixB_unitarity_h1}
\end{equation}
and
\begin{equation}
F_\ell(z)=
\sqrt{
\frac{3\rho}{
2\left[
1-J_0^2(d_{F_\ell})-J_1^2(d_{F_\ell})
\right]}
}\,
\frac{
J_1\left[
d_{F_\ell}\left(1-\frac{3}{4}\rho z\right)
\right]
}{
1-\frac{3}{4}\rho z
} \,.
\label{eq:appendixB_unitarity_Fl}
\end{equation}

The coupling of the scalar mode $F_\ell$ to two lightest massive gravitons is proportional to the constant $c_\ell$ defined as
\begin{equation}
c_\ell=
\int_0^{z_s}dz\,e^{-3A(z)}
F_\ell(z)
\left|
e^{A(z)}\frac{dh_1(z)}{dz}
\right|^2 \,. \label{eq:appendixB_unitarity_cl}
\end{equation}
The corresponding ratio $\frac{c_\ell}{c_1}$ can be written as
\begin{equation}
\frac{c_\ell}{c_1}
=
\sqrt{
\frac{
1-J_0^2(d_{F_1})-J_1^2(d_{F_1})
}{
1-J_0^2(d_{F_\ell})-J_1^2(d_{F_\ell})
}
}
\,
\frac{
\displaystyle
\int_0^1 dt\,t^{-2/3}
J_1(d_{F_\ell}t)J_1^2(d_{h_1}t)
}{
\displaystyle
\int_0^1 dt\,t^{-2/3}
J_1(d_{F_1}t)J_1^2(d_{h_1}t)
} \,. \label{eq:appendixB_unitarity_cl_ratio}
\end{equation}

The characteristic quantity from the scalar contribution to the unitarity sum rule of the amplitude $\mathcal M(h_1 h_1\to h_1 h_1)$ is defined then as~\cite{Bertucci:2026osd}
\begin{equation}
S_N\equiv
\sum_{\ell=1}^{N}
\left(
R_{\ell1}^2-\frac{4}{3}
\right)
\left(\frac{c_\ell}{c_1}\right)^2,
\qquad
R_{\ell1}=\frac{d_{F_\ell}}{d_{h_1}} \,.
\label{eq:appendixB_unitarity_SN}
\end{equation}
For the first two scalar modes,
\begin{equation}
S_2\simeq -0.68735.
\label{eq:appendixB_unitarity_S2}
\end{equation}
Including successively higher scalar modes, the sum converges numerically to
\begin{equation}
S_\infty\equiv\lim_{N\to\infty}S_N
\simeq -0.2966<0 \,. \label{eq:appendixB_unitarity_Sinf}
\end{equation}
Thus, the complete scalar tower gives a negative contribution, as required by the unitarity constraint, Ref.~\cite{Bertucci:2026osd}.

\bibliographystyle{JHEP}
\bibliography{biblio}

@article{Chivukula:2023qrt,
    author = "Chivukula, R. Sekhar and Gill, Joshua A. and Mohan, Kirtimaan A. and Sengupta, Dipan and Simmons, Elizabeth H. and Wang, Xing",
    title = "{Symmetries, spin-2 scattering amplitudes, and equivalence theorems in warped five-dimensional gravitational theories}",
    eprint = "2312.08576",
    archivePrefix = "arXiv",
    primaryClass = "hep-ph",
    reportNumber = "ADP23-28/T1237, MSUHEP-23-031",
    doi = "10.1103/PhysRevD.109.075016",
    journal = "Phys. Rev. D",
    volume = "109",
    number = "7",
    pages = "075016",
    year = "2024"
}

@article{Chivukula:2023sua,
    author = "Chivukula, R. Sekhar and Gill, Joshua A. and Mohan, Kirtimaan A. and Sengupta, Dipan and Simmons, Elizabeth H. and Wang, Xing",
    title = "{Scattering amplitudes of massive spin-2 Kaluza-Klein states with matter}",
    eprint = "2311.00770",
    archivePrefix = "arXiv",
    primaryClass = "hep-ph",
    reportNumber = "ADP-23-26/T1235, MSUHEP-23-029",
    doi = "10.1103/PhysRevD.109.015033",
    journal = "Phys. Rev. D",
    volume = "109",
    number = "1",
    pages = "015033",
    year = "2024"
}

@article{Chivukula:2022tla,
    author = "Chivukula, R. Sekhar and Foren, Dennis and Mohan, Kirtimaan A. and Sengupta, Dipan and Simmons, Elizabeth H.",
    title = "{Spin-2 Kaluza-Klein scattering in a stabilized warped background}",
    eprint = "2206.10628",
    archivePrefix = "arXiv",
    primaryClass = "hep-ph",
    reportNumber = "MSUHEP-22-021, ADP-22-20/T1191",
    doi = "10.1103/PhysRevD.107.035015",
    journal = "Phys. Rev. D",
    volume = "107",
    number = "3",
    pages = "035015",
    year = "2023"
}

@article{Chivukula:2021xod,
    author = "Chivukula, R. Sekhar and Foren, Dennis and Mohan, Kirtimaan A. and Sengupta, Dipan and Simmons, Elizabeth H.",
    title = "{Spin-2 Kaluza-Klein mode scattering in models with a massive radion}",
    eprint = "2104.08169",
    archivePrefix = "arXiv",
    primaryClass = "hep-ph",
    reportNumber = "MSUHEP-21-008",
    doi = "10.1103/PhysRevD.103.095024",
    journal = "Phys. Rev. D",
    volume = "103",
    number = "9",
    pages = "095024",
    year = "2021"
}

@article{Bellazzini:2013fga,
    author = "Bellazzini, Brando and Csaki, Csaba and Hubisz, Jay and Serra, Javi and Terning, John",
    title = "{A Naturally Light Dilaton and a Small Cosmological Constant}",
    eprint = "1305.3919",
    archivePrefix = "arXiv",
    primaryClass = "hep-th",
    doi = "10.1140/epjc/s10052-014-2790-x",
    journal = "Eur. Phys. J. C",
    volume = "74",
    pages = "2790",
    year = "2014"
}

@article{Piai:2026rst,
    author = "Piai, Maurizio and Rucinski, James",
    title = "{Light dilaton from top-down holographic confinement with magnetic fluxes}",
    eprint = "2602.14924",
    archivePrefix = "arXiv",
    primaryClass = "hep-th",
    doi = "10.1103/pdxc-3cjx",
    journal = "Phys. Rev. D",
    volume = "114",
    number = "2",
    pages = "026029",
    year = "2026"
}

@article{Elander:2025fpk,
    author = "Elander, Daniel and Faedo, Ant{\'o}n F. and Piai, Maurizio and Rodgers, Ronnie and Subils, Javier G.",
    title = "{Light dilaton near critical points in top-down holography}",
    eprint = "2502.19226",
    archivePrefix = "arXiv",
    primaryClass = "hep-th",
    reportNumber = "NORDITA 2025-009",
    doi = "10.1103/1wqc-vk7q",
    journal = "Phys. Rev. D",
    volume = "112",
    number = "12",
    pages = "126020",
    year = "2025"
}

@article{Elander:2020fmv,
    author = "Elander, Daniel and Piai, Maurizio and Roughley, John",
    title = "{Light dilaton in a metastable vacuum}",
    eprint = "2011.07049",
    archivePrefix = "arXiv",
    primaryClass = "hep-th",
    doi = "10.1103/PhysRevD.103.046009",
    journal = "Phys. Rev. D",
    volume = "103",
    number = "4",
    pages = "046009",
    year = "2021"
}

@article{Megias:2014iwa,
    author = "Meg\'{\i}as, Eugenio and Pujol\`as, Oriol",
    title = "{Naturally light dilatons from nearly marginal deformations}",
    eprint = "1401.4998",
    archivePrefix = "arXiv",
    primaryClass = "hep-th",
    reportNumber = "UAB-FT-751",
    doi = "10.1007/JHEP08(2014)081",
    journal = "JHEP",
    volume = "08",
    pages = "081",
    year = "2014"
}

@article{Fichet:2023dju,
    author = "Fichet, Sylvain and Meg\'{\i}as, Eugenio and Quir\'os, Mariano",
    title = "{Holographic fluids from 5D dilaton gravity}",
    eprint = "2311.14233",
    archivePrefix = "arXiv",
    primaryClass = "hep-th",
    doi = "10.1007/JHEP08(2024)077",
    journal = "JHEP",
    volume = "08",
    pages = "077",
    year = "2024"
}

@article{DeWolfe:1999cp,
    author = "DeWolfe, O. and Freedman, D. Z. and Gubser, S. S. and Karch, A.",
    title = "{Modeling the fifth-dimension with scalars and gravity}",
    eprint = "hep-th/9909134",
    archivePrefix = "arXiv",
    reportNumber = "HUTP-99-A048, MIT-CTP-2903",
    doi = "10.1103/PhysRevD.62.046008",
    journal = "Phys. Rev. D",
    volume = "62",
    pages = "046008",
    year = "2000"
}

@article{Antoniadis:2011qw,
    author = "Antoniadis, Ignatios and Arvanitaki, Asimina and Dimopoulos, Savas and Giveon, Amit",
    title = "{Phenomenology of TeV Little String Theory from Holography}",
    eprint = "1102.4043",
    archivePrefix = "arXiv",
    primaryClass = "hep-ph",
    reportNumber = "CERN-PH-TH-2011-024",
    doi = "10.1103/PhysRevLett.108.081602",
    journal = "Phys. Rev. Lett.",
    volume = "108",
    pages = "081602",
    year = "2012"
}

@article{Antoniadis:2001sw,
    author = "Antoniadis, Ignatios and Dimopoulos, Savas and Giveon, Amit",
    title = "{Little string theory at a TeV}",
    eprint = "hep-th/0103033",
    archivePrefix = "arXiv",
    reportNumber = "CERN-TH-2001-066, RI-07-00, ITP-01-04",
    doi = "10.1088/1126-6708/2001/05/055",
    journal = "JHEP",
    volume = "05",
    pages = "055",
    year = "2001"
}

@article{Aharony:1999ks,
    author = "Aharony, Ofer",
    editor = "Lechtenfeld, O. and Louis, J. and Lust, Dieter and Nicolai, H.",
    title = "{A Brief review of 'little string theories'}",
    eprint = "hep-th/9911147",
    archivePrefix = "arXiv",
    reportNumber = "RUNHETC-99-42",
    doi = "10.1088/0264-9381/17/5/302",
    journal = "Class. Quant. Grav.",
    volume = "17",
    pages = "929--938",
    year = "2000"
}

@article{Megias:2026wwp,
    author = "Meg\'{\i}as, Eugenio and Quir\'os, Mariano",
    title = "{Linear Dilaton in Cosmology and Particle Physics}",
    eprint = "2608.16772",
    archivePrefix = "arXiv",
    primaryClass = "hep-th",
    doi = "10.3390/universe12090287",
    journal = "Universe",
    volume = "12",
    number = "9",
    pages = "287",
    year = "2026"
}

@article{Megias:2021mgj,
    author = "Meg\'{\i}as, Eugenio and Quir\'os, Mariano",
    title = "{The Continuum Linear Dilaton}",
    eprint = "2104.10260",
    archivePrefix = "arXiv",
    primaryClass = "hep-ph",
    doi = "10.5506/APhysPolB.52.711",
    journal = "Acta Phys. Polon. B",
    volume = "52",
    number = "6-7",
    pages = "711",
    year = "2021"
}

@article{Gubser:2000nd,
    author = "Gubser, Steven S.",
    title = "{Curvature singularities: The Good, the bad, and the naked}",
    eprint = "hep-th/0002160",
    archivePrefix = "arXiv",
    reportNumber = "PUPT-1916",
    doi = "10.4310/ATMP.2000.v4.n3.a6",
    journal = "Adv. Theor. Math. Phys.",
    volume = "4",
    pages = "679--745",
    year = "2000"
}

@article{Goldberger:1999uk,
    author = "Goldberger, Walter D. and Wise, Mark B.",
    title = "{Modulus stabilization with bulk fields}",
    eprint = "hep-ph/9907447",
    archivePrefix = "arXiv",
    reportNumber = "CALT-68-2232",
    doi = "10.1103/PhysRevLett.83.4922",
    journal = "Phys. Rev. Lett.",
    volume = "83",
    pages = "4922--4925",
    year = "1999"
}

@article{Bertucci:2026osd,
    author = "Bertucci, Francesco and Bonifacio, James and Hinterbichler, Kurt",
    title = "{Moduli Bounds from Spin-2 Sum Rules}",
    eprint = "2608.05273",
    archivePrefix = "arXiv",
    primaryClass = "hep-th",
    month = "8",
    year = "2026"
}

@article{Mirbabayi:2026saz,
    author = "Mirbabayi, Mehrdad and Villadoro, Giovanni",
    title = "{How Heavy Can Moduli Be?}",
    eprint = "2603.09817",
    archivePrefix = "arXiv",
    primaryClass = "hep-th",
    month = "3",
    year = "2026"
}

@article{Antoniadis:1998ig,
    author = "Antoniadis, Ignatios and Arkani-Hamed, Nima and Dimopoulos, Savas and Dvali, G. R.",
    title = "{New dimensions at a millimeter to a Fermi and superstrings at a TeV}",
    eprint = "hep-ph/9804398",
    archivePrefix = "arXiv",
    reportNumber = "SLAC-PUB-7801, SU-ITP-98-28, CPTH-S608-0498, IC-98-39",
    doi = "10.1016/S0370-2693(98)00860-0",
    journal = "Phys. Lett. B",
    volume = "436",
    pages = "257--263",
    year = "1998"
}

@article{Arkani-Hamed:1998jmv,
    author = "Arkani-Hamed, Nima and Dimopoulos, Savas and Dvali, G. R.",
    title = "{The Hierarchy problem and new dimensions at a millimeter}",
    eprint = "hep-ph/9803315",
    archivePrefix = "arXiv",
    reportNumber = "SLAC-PUB-7769, SU-ITP-98-13",
    doi = "10.1016/S0370-2693(98)00466-3",
    journal = "Phys. Lett. B",
    volume = "429",
    pages = "263--272",
    year = "1998"
}

@article{Marchesano:2024gul,
    author = "Marchesano, Fernando and Shiu, Gary and Weigand, Timo",
    title = "{The Standard Model from String Theory: What Have We Learned?}",
    eprint = "2401.01939",
    archivePrefix = "arXiv",
    primaryClass = "hep-th",
    reportNumber = "IFT-UAM/CSIC-24-01, ZMP-HH/24-01",
    doi = "10.1146/annurev-nucl-102622-012235",
    journal = "Ann. Rev. Nucl. Part. Sci.",
    volume = "74",
    number = "1",
    pages = "113--140",
    year = "2024"
}

@article{Megias:2015ory,
    author = "Meg\'{\i}as, Eugenio and Pujol\`as, Oriol and Quir\'os, Mariano",
    title = "{On dilatons and the LHC diphoton excess}",
    eprint = "1512.06106",
    archivePrefix = "arXiv",
    primaryClass = "hep-ph",
    reportNumber = "MPP-2015-213, UAB-FT-768",
    doi = "10.1007/JHEP05(2016)137",
    journal = "JHEP",
    volume = "05",
    pages = "137",
    year = "2016"
}

@article{Fichet:2023xbu,
    author = "Fichet, Sylvain and Meg\'{\i}as, Eugenio and Quir\'os, Mariano",
    title = "{Holography of linear dilaton spacetimes from the bottom up}",
    eprint = "2309.02489",
    archivePrefix = "arXiv",
    primaryClass = "hep-th",
    doi = "10.1103/PhysRevD.109.106011",
    journal = "Phys. Rev. D",
    volume = "109",
    number = "10",
    pages = "106011",
    year = "2024"
}

@article{Barbosa:2024pyn,
    author = "Barbosa, Sergio and Fichet, Sylvain and Meg\'{\i}as, Eugenio and Quir\'os, Mariano",
    title = "{Entanglement entropy and thermal phase transitions from curvature singularities}",
    eprint = "2406.02899",
    archivePrefix = "arXiv",
    primaryClass = "hep-th",
    doi = "10.1007/JHEP04(2025)044",
    journal = "JHEP",
    volume = "04",
    pages = "044",
    year = "2025"
}

@article{Cabrer:2009we,
      author         = "Cabrer, Joan A. and von Gersdorff, Gero and Quir\'os,
                        Mariano",
      title          = "{Soft-Wall Stabilization}",
      journal        = "New J. Phys.",
      volume         = "12",
      year           = "2010",
      pages          = "075012",
      doi            = "10.1088/1367-2630/12/7/075012",
      eprint         = "0907.5361",
      archivePrefix  = "arXiv",
      primaryClass   = "hep-ph",
      reportNumber   = "CERN-PH-TH-2009-141",
      SLACcitation   = "%%CITATION = ARXIV:0907.5361;%%"
}

@article{Agashe:2007zd,
      author         = "Agashe, Kaustubh and Davoudiasl, Hooman and Perez, Gilad
                        and Soni, Amarjit",
      title          = "{Warped Gravitons at the LHC and Beyond}",
      journal        = "Phys. Rev.",
      volume         = "D76",
      year           = "2007",
      pages          = "036006",
      doi            = "10.1103/PhysRevD.76.036006",
      eprint         = "hep-ph/0701186",
      archivePrefix  = "arXiv",
      primaryClass   = "hep-ph",
      reportNumber   = "SU-4252-843, BNL-HET-07-3, YITP-SB-07-02",
      SLACcitation   = "%%CITATION = HEP-PH/0701186;%%"
}

@article{Aharony:1999ti,
      author         = "Aharony, Ofer and Gubser, Steven S. and Maldacena, Juan
                        Martin and Ooguri, Hirosi and Oz, Yaron",
      title          = "{Large N field theories, string theory and gravity}",
      journal        = "Phys. Rept.",
      volume         = "323",
      year           = "2000",
      pages          = "183-386",
      doi            = "10.1016/S0370-1573(99)00083-6",
      eprint         = "hep-th/9905111",
      archivePrefix  = "arXiv",
      primaryClass   = "hep-th",
      reportNumber   = "CERN-TH-99-122, HUTP-99-A027, LBNL-43113, RU-99-18,
                        UCB-PTH-99-16, LBL-43113",
      SLACcitation   = "%%CITATION = HEP-TH/9905111;%%"
}

@article{Randall:1999ee,
      author         = "Randall, Lisa and Sundrum, Raman",
      title          = "{A Large mass hierarchy from a small extra dimension}",
      journal        = "Phys. Rev. Lett.",
      volume         = "83",
      year           = "1999",
      pages          = "3370-3373",
      doi            = "10.1103/PhysRevLett.83.3370",
      eprint         = "hep-ph/9905221",
      archivePrefix  = "arXiv",
      primaryClass   = "hep-ph",
      reportNumber   = "MIT-CTP-2860, PUPT-1860, BUHEP-99-9",
      SLACcitation   = "%%CITATION = HEP-PH/9905221;%%"
}

@article{Randall:1999vf,
      author         = "Randall, Lisa and Sundrum, Raman",
      title          = "{An Alternative to compactification}",
      journal        = "Phys. Rev. Lett.",
      volume         = "83",
      year           = "1999",
      pages          = "4690-4693",
      doi            = "10.1103/PhysRevLett.83.4690",
      eprint         = "hep-th/9906064",
      archivePrefix  = "arXiv",
      primaryClass   = "hep-th",
      reportNumber   = "MIT-CTP-2874, PUPT-1867, BUHEP-99-13",
      SLACcitation   = "%%CITATION = HEP-TH/9906064;%%"
}

@article{Pomarol:2019aae,
    author = "Pomarol, Alex and Pujol\`as, Oriol and Salas, Lindber",
    title = "{Holographic conformal transition and light scalars}",
    eprint = "1905.02653",
    archivePrefix = "arXiv",
    primaryClass = "hep-th",
    doi = "10.1007/JHEP10(2019)202",
    journal = "JHEP",
    volume = "10",
    pages = "202",
    year = "2019"
}

@article{Pomarol:2023xcc,
    author = "Pomarol, Alex and Salas, Lindber",
    title = "{Exploring the conformal transition from above and below}",
    eprint = "2312.08332",
    archivePrefix = "arXiv",
    primaryClass = "hep-ph",
    doi = "10.1007/JHEP08(2024)149",
    journal = "JHEP",
    volume = "08",
    pages = "149",
    year = "2024"
}

@article{Kato:1995,
    author = "Kato, Tosio",
    title = "{Perturbation Theory for Linear Operators}",
    journal = "Springer, Classics in Mathematics",
    volume = "2nd ed.",
    pages = "1--619",
    doi = "10.1007/978-3-642-66282-9",
    year = "1995"
}

@article{Teschl:2014,
    author = "Teschl, Gerald",
    title = "{Mathematical Methods in Quantum Mechanics: With Applications to Schrödinger Operators}",
    journal = "Graduate Studies in Mathematics",
    volume = "157",
    pages = "1--358",
    year = "2014"
}

@article{Watson:1944,
    author = "Watson, G. N.",
    title = "{A Treatise on the Theory of Bessel Functions}",
    journal = "Cambridge University Press",
    volume = "2nd ed.",
    pages = "1--804",
    year = "1944"
}

@article{Olver:1951,
    author = "Olver, F. W. J.",
    title = "{A Further Method for the Evaluation of Zeros of Bessel Functions and Some New Asymptotic Expansions for Zeros of Functions of Large Order}",
    journal = "Proc. Cambridge Phil. Soc.",
    volume = "47",
    pages = "699--712",
    year = "1951"
}

\end{document}